
%
\expandafter\ifx\csname phyzzx\endcsname\relax
 \message{It is better to use PHYZZX format than to
          \string\input\space PHYZZX}\else
 \wlog{PHYZZX macros are already loaded and are not
          \string\input\space again}%
 \endinput \fi
\catcode`\@=11 
\let\rel@x=\relax
\let\n@expand=\relax
\def\pr@tect{\let\n@expand=\noexpand}
\let\protect=\pr@tect
\let\gl@bal=\global
%
%
%
\newfam\cpfam
\newdimen\b@gheight             \b@gheight=12pt
\newcount\f@ntkey               \f@ntkey=0
\def\f@m{\afterassignment\samef@nt\f@ntkey=}
\def\samef@nt{\fam=\f@ntkey\the\textfont\f@ntkey\rel@x}
\def\setstr@t{\setbox\strutbox=\hbox{\vrule height 0.85\b@gheight
                                depth 0.35\b@gheight width\z@ }}
%

%
%
%
%
\font\seventeenrm =cmr17
\font\fourteenrm  =cmr12 scaled\magstep1
\font\twelverm    =cmr12
\font\ninerm      =cmr9
\font\sixrm       =cmr6

\font\fourteenbf  =cmbx12 scaled\magstep1
\font\twelvebf    =cmbx12
\font\ninebf      =cmbx9
\font\sixbf       =cmbx6
\font\seventeeni  =cmmi12 scaled\magstep2    \skewchar\seventeeni='177
\font\fourteeni   =cmmi12 scaled\magstep1     \skewchar\fourteeni='177
\font\twelvei     =cmmi12                       \skewchar\twelvei='177
\font\ninei       =cmmi9                          \skewchar\ninei='177
\font\sixi        =cmmi6                           \skewchar\sixi='177
\font\seventeensy =cmsy10 scaled\magstep3    \skewchar\seventeensy='60
\font\fourteensy  =cmsy10 scaled\magstep2     \skewchar\fourteensy='60
\font\twelvesy    =cmsy10 scaled\magstep1       \skewchar\twelvesy='60
\font\ninesy      =cmsy9                          \skewchar\ninesy='60
\font\sixsy       =cmsy6                           \skewchar\sixsy='60

\font\fourteenex  =cmex10 scaled\magstep2
\font\twelveex    =cmex10 scaled\magstep1

\font\fourteensl  =cmsl12 scaled\magstep1
\font\twelvesl    =cmsl12
\font\ninesl      =cmsl9

\font\fourteenit  =cmti12 scaled\magstep1
\font\twelveit    =cmti12
\font\nineit      =cmti9
\font\fourteentt  =cmtt12 scaled\magstep1
\font\twelvett    =cmtt12
\font\fourteencp  =cmcsc10 scaled\magstep2
\font\twelvecp    =cmcsc10 scaled\magstep1
\font\tencp       =cmcsc10
%
%
\def\fourteenf@nts{\relax
    \textfont0=\fourteenrm          \scriptfont0=\tenrm
      \scriptscriptfont0=\sevenrm
    \textfont1=\fourteeni           \scriptfont1=\teni
      \scriptscriptfont1=\seveni
    \textfont2=\fourteensy          \scriptfont2=\tensy
      \scriptscriptfont2=\sevensy
    \textfont3=\fourteenex          \scriptfont3=\twelveex
      \scriptscriptfont3=\tenex
    \textfont\itfam=\fourteenit     \scriptfont\itfam=\tenit
    \textfont\slfam=\fourteensl     \scriptfont\slfam=\tensl
    \textfont\bffam=\fourteenbf     \scriptfont\bffam=\tenbf
      \scriptscriptfont\bffam=\sevenbf
    \textfont\ttfam=\fourteentt
    \textfont\cpfam=\fourteencp }
\def\twelvef@nts{\relax
    \textfont0=\twelverm          \scriptfont0=\ninerm
      \scriptscriptfont0=\sixrm
    \textfont1=\twelvei           \scriptfont1=\ninei
      \scriptscriptfont1=\sixi
    \textfont2=\twelvesy           \scriptfont2=\ninesy
      \scriptscriptfont2=\sixsy
    \textfont3=\twelveex          \scriptfont3=\tenex
      \scriptscriptfont3=\tenex
    \textfont\itfam=\twelveit     \scriptfont\itfam=\nineit
    \textfont\slfam=\twelvesl     \scriptfont\slfam=\ninesl
    \textfont\bffam=\twelvebf     \scriptfont\bffam=\ninebf
      \scriptscriptfont\bffam=\sixbf
    \textfont\ttfam=\twelvett
    \textfont\cpfam=\twelvecp }
\def\tenf@nts{\relax
    \textfont0=\tenrm          \scriptfont0=\sevenrm
      \scriptscriptfont0=\fiverm
    \textfont1=\teni           \scriptfont1=\seveni
      \scriptscriptfont1=\fivei
    \textfont2=\tensy          \scriptfont2=\sevensy
      \scriptscriptfont2=\fivesy
    \textfont3=\tenex          \scriptfont3=\tenex
      \scriptscriptfont3=\tenex
    \textfont\itfam=\tenit     \scriptfont\itfam=\seveni  
    \textfont\slfam=\tensl     \scriptfont\slfam=\sevenrm 
    \textfont\bffam=\tenbf     \scriptfont\bffam=\sevenbf
      \scriptscriptfont\bffam=\fivebf
    \textfont\ttfam=\tentt
    \textfont\cpfam=\tencp }
%
%

%
\def\rm{\n@expand\f@m0 }
\def\mit{\n@expand\f@m1 }         \let\oldstyle=\mit
\def\cal{\n@expand\f@m2}
\def\it{\n@expand\f@m\itfam}
\def\sl{\n@expand\f@m\slfam}
\def\bf{\n@expand\f@m\bffam}
\def\tt{\n@expand\f@m\ttfam}
\def\caps{\n@expand\f@m\cpfam}    \let\cp=\caps
\def\em@{\rel@x\ifnum\f@ntkey=0\it\else
        \ifnum\f@ntkey=\bffam\it\else\rm\fi \fi }
\def\em{\n@expand\em@}
\def\fourteenpoint{\fourteenf@nts \samef@nt \b@gheight=14pt \setstr@t }
\def\twelvepoint{\twelvef@nts \samef@nt \b@gheight=12pt \setstr@t }
\def\tenpoint{\tenf@nts \samef@nt \b@gheight=10pt \setstr@t }
\normalbaselineskip = 20pt plus 0.2pt minus 0.1pt
\normallineskip = 1.5pt plus 0.1pt minus 0.1pt
\normallineskiplimit = 1.5pt
\newskip\normaldisplayskip
\normaldisplayskip = 20pt plus 5pt minus 10pt
\newskip\normaldispshortskip
\normaldispshortskip = 6pt plus 5pt
\newskip\normalparskip
\normalparskip = 6pt plus 2pt minus 1pt
\newskip\skipregister
\skipregister = 5pt plus 2pt minus 1.5pt
\newif\ifsingl@
\newif\ifdoubl@
\newif\iftwelv@  \twelv@true
\def\singlespace{\singl@true\doubl@false\spaces@t}
\def\doublespace{\singl@false\doubl@true\spaces@t}
\def\normalspace{\singl@false\doubl@false\spaces@t}
\def\Tenpoint{\tenpoint\twelv@false\spaces@t}
\def\Twelvepoint{\twelvepoint\twelv@true\spaces@t}
\def\spaces@t{\rel@x
      \iftwelv@ \ifsingl@\subspaces@t3:4;\else\subspaces@t1:1;\fi
       \else \ifsingl@\subspaces@t3:5;\else\subspaces@t4:5;\fi \fi
      \ifdoubl@ \multiply\baselineskip by 5
         \divide\baselineskip by 4 \fi }
\def\subspaces@t#1:#2;{
      \baselineskip = \normalbaselineskip
      \multiply\baselineskip by #1 \divide\baselineskip by #2
      \lineskip = \normallineskip
      \multiply\lineskip by #1 \divide\lineskip by #2
      \lineskiplimit = \normallineskiplimit
      \multiply\lineskiplimit by #1 \divide\lineskiplimit by #2
      \parskip = \normalparskip
      \multiply\parskip by #1 \divide\parskip by #2
      \abovedisplayskip = \normaldisplayskip
      \multiply\abovedisplayskip by #1 \divide\abovedisplayskip by #2
      \belowdisplayskip = \abovedisplayskip
      \abovedisplayshortskip = \normaldispshortskip
      \multiply\abovedisplayshortskip by #1
        \divide\abovedisplayshortskip by #2
      \belowdisplayshortskip = \abovedisplayshortskip
      \advance\belowdisplayshortskip by \belowdisplayskip
      \divide\belowdisplayshortskip by 2
      \smallskipamount = \skipregister
      \multiply\smallskipamount by #1 \divide\smallskipamount by #2
      \medskipamount = \smallskipamount \multiply\medskipamount by 2
      \bigskipamount = \smallskipamount \multiply\bigskipamount by 4 }
\def\normalbaselines{ \baselineskip=\normalbaselineskip
   \lineskip=\normallineskip \lineskiplimit=\normallineskip
   \iftwelv@\else \multiply\baselineskip by 4 \divide\baselineskip by 5
     \multiply\lineskiplimit by 4 \divide\lineskiplimit by 5
     \multiply\lineskip by 4 \divide\lineskip by 5 \fi }
\Twelvepoint  
\interlinepenalty=50
\interfootnotelinepenalty=5000
\predisplaypenalty=9000
\postdisplaypenalty=500
\hfuzz=1pt
\vfuzz=0.2pt
\newdimen\HOFFSET  \HOFFSET=0pt
\newdimen\VOFFSET  \VOFFSET=0pt
\newdimen\HSWING   \HSWING=0pt
\dimen\footins=8in
%
%
%
\newskip\pagebottomfiller
\pagebottomfiller=\z@ plus \z@ minus \z@
\def\pagecontents{
   \ifvoid\topins\else\unvbox\topins\vskip\skip\topins\fi
   \dimen@ = \dp255 \unvbox255
   \vskip\pagebottomfiller
   \ifvoid\footins\else\vskip\skip\footins\footrule\unvbox\footins\fi
   \ifr@ggedbottom \kern-\dimen@ \vfil \fi }
\def\makeheadline{\vbox to 0pt{ \skip@=\topskip
      \advance\skip@ by -12pt \advance\skip@ by -2\normalbaselineskip
      \vskip\skip@ \line{\vbox to 12pt{}\the\headline} \vss
      }\nointerlineskip}
\def\makefootline{\baselineskip = 1.5\normalbaselineskip
                 \line{\the\footline}}
\newif\iffrontpage
\newif\ifp@genum
\def\nopagenumbers{\p@genumfalse}
\def\pagenumbers{\p@genumtrue}
\pagenumbers
\newtoks\paperheadline
\newtoks\paperfootline
\newtoks\letterheadline
\newtoks\letterfootline
\newtoks\letterinfo
\newtoks\date
\paperheadline={\hfil}
\paperfootline={\hss\iffrontpage\else\ifp@genum\tenrm\folio\hss\fi\fi}
\letterheadline{\iffrontpage \hfil \else
    \rm \ifp@genum page~~\folio\fi \hfil\the\date \fi}
\letterfootline={\iffrontpage\the\letterinfo\else\hfil\fi}
\letterinfo={\hfil}
\def\monthname{\rel@x\ifcase\month 0/\or January\or February\or
   March\or April\or May\or June\or July\or August\or September\or
   October\or November\or December\else\number\month/\fi}
\def\today{\monthname~\number\day, \number\year}
\date={\today}
\headline=\paperheadline 
\footline=\paperfootline 
\countdef\pageno=1      \countdef\pagen@=0
\countdef\pagenumber=1  \pagenumber=1
\def\advancepageno{\gl@bal\advance\pagen@ by 1
   \ifnum\pagenumber<0 \gl@bal\advance\pagenumber by -1
    \else\gl@bal\advance\pagenumber by 1 \fi
    \gl@bal\frontpagefalse  \swing@ }
\def\folio{\ifnum\pagenumber<0 \romannumeral-\pagenumber
           \else \number\pagenumber \fi }
\def\swing@{\ifodd\pagenumber \gl@bal\advance\hoffset by -\HSWING
             \else \gl@bal\advance\hoffset by \HSWING \fi }
\def\footrule{\dimen@=\prevdepth\nointerlineskip
   \vbox to 0pt{\vskip -0.25\baselineskip \hrule width 0.35\hsize \vss}
   \prevdepth=\dimen@ }
\let\footnotespecial=\rel@x
\newdimen\footindent
\footindent=24pt
\def\Textindent#1{\noindent\llap{#1\enspace}\ignorespaces}
\def\Vfootnote#1{\insert\footins\bgroup
   \interlinepenalty=\interfootnotelinepenalty \floatingpenalty=20000
   \singl@true\doubl@false\Tenpoint
   \splittopskip=\ht\strutbox \boxmaxdepth=\dp\strutbox
   \leftskip=\footindent \rightskip=\z@skip
   \parindent=0.5\footindent \parfillskip=0pt plus 1fil
   \spaceskip=\z@skip \xspaceskip=\z@skip \footnotespecial
   \Textindent{#1}\footstrut\futurelet\next\fo@t}

\def\vfootnote#1{\Vfootnote{${#1}$}}
\def\footnote#1{\attach{#1}\vfootnote{#1}}

\let\footsymbol=\star
\newcount\lastf@@t           \lastf@@t=-1
\newcount\footsymbolcount    \footsymbolcount=0
\newif\ifPhysRev
\def\bumpfootsymbolcount{\rel@x
   \iffrontpage \bumpfootsymbolpos \else \advance\lastf@@t by 1
     \ifPhysRev \bumpfootsymbolneg \else \bumpfootsymbolpos \fi \fi
   \gl@bal\lastf@@t=\pagen@ }
\def\bumpfootsymbolpos{\ifnum\footsymbolcount <0
                            \gl@bal\footsymbolcount =0 \fi
    \ifnum\lastf@@t<\pagen@ \gl@bal\footsymbolcount=0
     \else \gl@bal\advance\footsymbolcount by 1 \fi }
\def\bumpfootsymbolneg{\ifnum\footsymbolcount >0
             \gl@bal\footsymbolcount =0 \fi
         \gl@bal\advance\footsymbolcount by -1 }
\def\fd@f#1 {\xdef\footsymbol{\mathchar"#1 }}
\def\generatefootsymbol{\ifcase\footsymbolcount \fd@f 13F \or \fd@f 279
        \or \fd@f 27A \or \fd@f 278 \or \fd@f 27B \else
        \ifnum\footsymbolcount <0 \fd@f{023 \number-\footsymbolcount }
         \else \fd@f 203 {\loop \ifnum\footsymbolcount >5
                \fd@f{203 \footsymbol } \advance\footsymbolcount by -1
                \repeat }\fi \fi }

\def\nonfrenchspacing{\sfcode`\.=3001 \sfcode`\!=3000 \sfcode`\?=3000
        \sfcode`\:=2000 \sfcode`\;=1500 \sfcode`\,=1251 }
\nonfrenchspacing
\newdimen\d@twidth
{\setbox0=\hbox{s.} \gl@bal\d@twidth=\wd0 \setbox0=\hbox{s}
        \gl@bal\advance\d@twidth by -\wd0 }
\def\removehglue{\loop \unskip \ifdim\lastskip >\z@ \repeat }
\def\roll@ver#1{\removehglue \nobreak \count255 =\spacefactor \dimen@=\z@
        \ifnum\count255 =3001 \dimen@=\d@twidth \fi
        \ifnum\count255 =1251 \dimen@=\d@twidth \fi
    \iftwelv@ \kern-\dimen@ \else \kern-0.83\dimen@ \fi
   #1\spacefactor=\count255 }
\def\step@ver#1{\rel@x \ifmmode #1\else \ifhmode
        \roll@ver{${}#1$}\else {\setbox0=\hbox{${}#1$}}\fi\fi }
\def\attach#1{\step@ver{\strut^{\mkern 2mu #1} }}
%
%
%
\newcount\chapternumber      \chapternumber=0
\newcount\sectionnumber      \sectionnumber=0
\newcount\equanumber         \equanumber=0
\let\chapterlabel=\rel@x
\let\sectionlabel=\rel@x
\newtoks\chapterstyle        \chapterstyle={\Number}
\newtoks\sectionstyle        \sectionstyle={\Number}
\newskip\chapterskip         \chapterskip=\bigskipamount
\newskip\sectionskip         \sectionskip=\medskipamount
\newskip\headskip            \headskip=8pt plus 3pt minus 3pt
\newdimen\chapterminspace    \chapterminspace=15pc
\newdimen\sectionminspace    \sectionminspace=10pc
\newdimen\referenceminspace  \referenceminspace=20pc
\newif\ifcn@                 \cn@true
\newif\ifcn@@                \cn@@false
\def\numberedchapters{\cn@true}
\def\unnumberedchapters{\cn@false\sequentialequations}
\def\chapterreset{\gl@bal\advance\chapternumber by 1
   \ifnum\equanumber<0 \else\gl@bal\equanumber=0\fi
   \sectionnumber=0 \let\sectionlabel=\rel@x
   \ifcn@ \gl@bal\cn@@true {\pr@tect
       \xdef\chapterlabel{\the\chapterstyle{\the\chapternumber}}}%
    \else \gl@bal\cn@@false \gdef\chapterlabel{\rel@x}\fi }
\def\@alpha#1{\count255='140 \advance\count255 by #1\char\count255}
 \def\alphabetic{\n@expand\@alpha}
\def\@Alpha#1{\count255='100 \advance\count255 by #1\char\count255}
 \def\Alphabetic{\n@expand\@Alpha}
\def\@Roman#1{\uppercase\expandafter{\romannumeral #1}}
 \def\Roman{\n@expand\@Roman}
\def\@roman#1{\romannumeral #1}    \def\roman{\n@expand\@roman}
\def\@number#1{\number #1}         \def\Number{\n@expand\@number}
\def\BLANK#1{\rel@x}               
\def\titleparagraphs{\interlinepenalty=9999
     \leftskip=0.03\hsize plus 0.22\hsize minus 0.03\hsize
     \rightskip=\leftskip \parfillskip=0pt
     \hyphenpenalty=9000 \exhyphenpenalty=9000
     \tolerance=9999 \pretolerance=9000
     \spaceskip=0.333em \xspaceskip=0.5em }
\def\titlestyle#1{\par\begingroup \titleparagraphs
     \iftwelv@\fourteenpoint\else\twelvepoint\fi
   \noindent #1\par\endgroup }
\def\spacecheck#1{\dimen@=\pagegoal\advance\dimen@ by -\pagetotal
   \ifdim\dimen@<#1 \ifdim\dimen@>0pt \vfil\break \fi\fi}
\def\chapter#1{\par \penalty-300 \vskip\chapterskip
   \spacecheck\chapterminspace
   \chapterreset \titlestyle{\ifcn@@\chapterlabel.~\fi #1}
   \nobreak\vskip\headskip \penalty 30000
   {\pr@tect\wlog{\string\chapter\space \chapterlabel}} }

\def\section#1{\par \ifnum\lastpenalty=30000\else
   \penalty-200\vskip\sectionskip \spacecheck\sectionminspace\fi
   \gl@bal\advance\sectionnumber by 1
   {\pr@tect
   \xdef\sectionlabel{\ifcn@@ \chapterlabel.\fi
       \the\sectionstyle{\the\sectionnumber}}%
   \wlog{\string\section\space \sectionlabel}}%
   \noindent {\caps\enspace\sectionlabel.~~#1}\par
   \nobreak\vskip\headskip \penalty 30000 }
\def\subsection#1{\par
   \ifnum\the\lastpenalty=30000\else \penalty-100\smallskip \fi
   \noindent\undertext{#1}\enspace \vadjust{\penalty5000}}

\def\undertext#1{\vtop{\hbox{#1}\kern 1pt \hrule}}
\def\APPENDIX#1#2{\par\penalty-300\vskip\chapterskip
   \spacecheck\chapterminspace \chapterreset \xdef\chapterlabel{#1}
   \titlestyle{APPENDIX #2} \nobreak\vskip\headskip \penalty 30000
   \wlog{\string\Appendix~\chapterlabel} }
\def\Appendix#1{\APPENDIX{#1}{#1}}
\def\appendix{\APPENDIX{A}{}}
%
%
%
%
\def\eq{\eqname\?}
\def\eqn{\eqno\eqname}

\def\eqinsert#1{\noalign{\dimen@=\prevdepth \nointerlineskip
   \setbox0=\hbox to\displaywidth{\hfil #1}
   \vbox to 0pt{\kern 0.5\baselineskip\hbox{$\!\box0\!$}\vss}
   \prevdepth=\dimen@}}
%

%
%
\def\GENITEM#1;#2{\par \hangafter=0 \hangindent=#1
    \Textindent{$ #2 $}\ignorespaces}
\outer\def\newitem#1=#2;{\gdef#1{\GENITEM #2;}}

\newdimen\itemsize                \itemsize=30pt
\newitem\item=1\itemsize;
\newitem\sitem=1.75\itemsize;     
\newitem\ssitem=2.5\itemsize;     
\outer\def\newlist#1=#2&#3&#4;{\toks0={#2}\toks1={#3}%
   \count255=\escapechar \escapechar=-1
   \alloc@0\list\countdef\insc@unt\listcount     \listcount=0
   \edef#1{\par
      \countdef\listcount=\the\allocationnumber
      \advance\listcount by 1
      \hangafter=0 \hangindent=#4
      \Textindent{\the\toks0{\listcount}\the\toks1}}
   \expandafter\expandafter\expandafter
    \edef\c@t#1{begin}{\par
      \countdef\listcount=\the\allocationnumber \listcount=1
      \hangafter=0 \hangindent=#4
      \Textindent{\the\toks0{\listcount}\the\toks1}}
   \expandafter\expandafter\expandafter
    \edef\c@t#1{con}{\par \hangafter=0 \hangindent=#4 \noindent}
   \escapechar=\count255}
\def\c@t#1#2{\csname\string#1#2\endcsname}
\newlist\point=\Number&.&1.0\itemsize;
\newlist\subpoint=(\alphabetic&)&1.75\itemsize;
\newlist\subsubpoint=(\roman&)&2.5\itemsize;
%

%
%
%
%
\newcount\referencecount     \referencecount=0
\newcount\lastrefsbegincount \lastrefsbegincount=0
\newif\ifreferenceopen       \newwrite\referencewrite
\newdimen\refindent          \refindent=30pt
\def\normalrefmark#1{\attach{\scriptscriptstyle [ #1 ] }}
\let\PRrefmark=\attach
\def\NPrefmark#1{\step@ver{{\;[#1]}}}
\def\refmark#1{\rel@x\ifPhysRev\PRrefmark{#1}\else\normalrefmark{#1}\fi}
\def\refend@{\refmark{\number\referencecount}}
\def\refend{\refend@{}\space }
\def\refsend{\refmark{\count255=\referencecount
   \advance\count255 by-\lastrefsbegincount
   \ifcase\count255 \number\referencecount
   \or \number\lastrefsbegincount,\number\referencecount
   \else \number\lastrefsbegincount-\number\referencecount \fi}\space }
\def\REFNUM#1{\rel@x \gl@bal\advance\referencecount by 1
    \xdef#1{\the\referencecount }}
\def\Refnum#1{\REFNUM #1\refend@ } 
\def\REF#1{\REFNUM #1\R@FWRITE\ignorespaces}
\def\Ref#1{\Refnum #1\REFWRITE }
\def\ref{\Ref\?}
\def\REFS#1{\REFNUM #1\gl@bal\lastrefsbegincount=\referencecount
    \REFWRITE }

\def\r@fitem#1{\par \hangafter=0 \hangindent=\refindent \Textindent{#1}}
\def\refitem#1{\r@fitem{#1.}}
\def\NPrefitem#1{\r@fitem{[#1]}}
\def\NPrefs{\let\refmark=\NPrefmark \let\refitem=NPrefitem}
\def\REFWRITE{\R@FWRITE\rel@x }
\def\R@FWRITE#1{\ifreferenceopen \else \gl@bal\referenceopentrue
     \immediate\openout\referencewrite=\jobname.refs
     \toks@={\begingroup \refoutspecials \catcode`\^^M=10 }%
     \immediate\write\referencewrite{\the\toks@}\fi
    \immediate\write\referencewrite{\noexpand\refitem %
                                    {\the\referencecount}}%
    \p@rse@ndwrite \referencewrite #1}
\begingroup
 \catcode`\^^M=\active \let^^M=\relax %
 \gdef\p@rse@ndwrite#1#2{\begingroup \catcode`\^^M=12 \newlinechar=`\^^M%
         \chardef\rw@write=#1\sc@nlines#2}%
 \gdef\sc@nlines#1#2{\sc@n@line \g@rbage #2^^M\endsc@n \endgroup #1}%
 \gdef\sc@n@line#1^^M{\expandafter\toks@\expandafter{\deg@rbage #1}%
         \immediate\write\rw@write{\the\toks@}%
         \futurelet\n@xt \sc@ntest }%
\endgroup
\def\sc@ntest{\ifx\n@xt\endsc@n \let\n@xt=\rel@x
       \else \let\n@xt=\sc@n@notherline \fi \n@xt }
\def\sc@n@notherline{\sc@n@line \g@rbage }
\def\deg@rbage#1{}
\let\g@rbage=\relax    \let\endsc@n=\relax
\def\refout{\par\penalty-400\vskip\chapterskip
   \spacecheck\referenceminspace
   \ifreferenceopen \Closeout\referencewrite \referenceopenfalse \fi
   \line{\fourteenrm\hfil REFERENCES\hfil}\vskip\headskip
   \input \jobname.refs
   }
\def\refoutspecials{\sfcode`\.=1000 \interlinepenalty=1000
         \rightskip=\z@ plus 1em minus \z@ }
\def\Closeout#1{\toks0={\par\endgroup}\immediate\write#1{\the\toks0}%
   \immediate\closeout#1}
%
%
\newcount\figurecount     \figurecount=0
\newcount\tablecount      \tablecount=0
\newif\iffigureopen       \newwrite\figurewrite
\newif\iftableopen        \newwrite\tablewrite
\def\FIGNUM#1{\rel@x \gl@bal\advance\figurecount by 1
    \xdef#1{\the\figurecount}}
\def\FIGURE#1{\FIGNUM #1\F@GWRITE\ignorespaces }
\let\FIG=\FIGURE

\def\figitem#1{\r@fitem{#1)}}
\def\FIGWRITE{\F@GWRITE\rel@x }
\def\TABNUM#1{\rel@x \gl@bal\advance\tablecount by 1
    \xdef#1{\the\tablecount}}
\def\TABLE#1{\TABNUM #1\T@BWRITE\ignorespaces }
\def\Table{\TABNUM\?Table~\?\TABWRITE }
\def\tabitem#1{\r@fitem{#1:}}
\def\TABWRITE{\T@BWRITE\rel@x }
\def\F@GWRITE#1{\iffigureopen \else \gl@bal\figureopentrue
     \immediate\openout\figurewrite=\jobname.figs
     \toks@={\begingroup \catcode`\^^M=10 }%
     \immediate\write\figurewrite{\the\toks@}\fi
    \immediate\write\figurewrite{\noexpand\figitem %
                                 {\the\figurecount}}%
    \p@rse@ndwrite \figurewrite #1}
\def\T@BWRITE#1{\iftableopen \else \gl@bal\tableopentrue
     \immediate\openout\tablewrite=\jobname.tabs
     \toks@={\begingroup \catcode`\^^M=10 }%
     \immediate\write\tablewrite{\the\toks@}\fi
    \immediate\write\tablewrite{\noexpand\tabitem %
                                 {\the\tablecount}}%
    \p@rse@ndwrite \tablewrite #1}
\def\figout{\par\penalty-400
   \vskip\chapterskip\spacecheck\referenceminspace
   \iffigureopen \Closeout\figurewrite \figureopenfalse \fi
   \line{\fourteenrm\hfil FIGURE CAPTIONS\hfil}\vskip\headskip
   \input \jobname.figs
   }
\def\tabout{\par\penalty-400
   \vskip\chapterskip\spacecheck\referenceminspace
   \iftableopen \Closeout\tablewrite \tableopenfalse \fi
   \line{\fourteenrm\hfil TABLE CAPTIONS\hfil}\vskip\headskip
   \input \jobname.tabs
   }
%
%
%
\newbox\picturebox
\def\p@cht{\ht\picturebox }
\def\p@cwd{\wd\picturebox }
\def\p@cdp{\dp\picturebox }
\newdimen\xshift
\newdimen\yshift
\newdimen\captionwidth
\newskip\captionskip
\captionskip=15pt plus 5pt minus 3pt
\def\fullwidth{\captionwidth=\hsize }
\newtoks\Caption
\newif\ifcaptioned
\newif\ifselfcaptioned
\def\caption{\captionedtrue \Caption }
\newcount\linesabove
\newif\iffileexists
\newtoks\picfilename
\def\fil@#1 {\fileexiststrue \picfilename={#1}}
\def\file#1{\if=#1\let\n@xt=\fil@ \else \def\n@xt{\fil@ #1}\fi \n@xt }
\def\pl@t{\begingroup \pr@tect
    \setbox\picturebox=\hbox{}\fileexistsfalse
    \let\height=\p@cht \let\width=\p@cwd \let\depth=\p@cdp
    \xshift=\z@ \yshift=\z@ \captionwidth=\z@
    \Caption={}\captionedfalse
    \linesabove =0 \picturedefault }
\def\plot{\pl@t \selfcaptionedfalse }
\def\Picture#1{\gl@bal\advance\figurecount by 1
    \xdef#1{\the\figurecount}\pl@t \selfcaptionedtrue }

\def\s@vepicture{\iffileexists \parsefilename \redopicturebox \fi
   \ifdim\captionwidth>\z@ \else \captionwidth=\p@cwd \fi
   \xdef\lastpicture{%
      \iffileexists%
         \setbox0=\hbox{\raise\the\yshift \vbox{%
              \moveright\the\xshift\hbox{\picturedefinition}}}%
      \else%
         \setbox0=\hbox{}%
      \fi
      \ht0=\the\p@cht \wd0=\the\p@cwd \dp0=\the\p@cdp
      \vbox{\hsize=\the\captionwidth%
            \line{\hss\box0 \hss }%
            \ifcaptioned%
               \vskip\the\captionskip \noexpand\Tenpoint
               \ifselfcaptioned%
                   Figure~\the\figurecount.\enspace%
               \fi%
               \the\Caption%
           \fi%
           }%
      }%
      \endgroup%
}
\let\endpicture=\s@vepicture
\def\savepicture#1{\s@vepicture \global\let#1=\lastpicture }
\def\displaypicture{\fullwidth \s@vepicture $$\lastpicture $${}}
\def\toppicture{\fullwidth \s@vepicture \topinsert
    \lastpicture \medskip \endinsert }
\def\midpicture{\fullwidth \s@vepicture \midinsert
    \lastpicture \endinsert }
%
%
\def\leftpicture{\pres@tpicture
    \dimen@i=\hsize \advance\dimen@i by -\dimen@ii
    \setbox\picturebox=\hbox to \hsize {\box0 \hss }%
    \wr@paround }
\def\rightpicture{\pres@tpicture
    \dimen@i=\z@
    \setbox\picturebox=\hbox to \hsize {\hss \box0 }%
    \wr@paround }
\def\pres@tpicture{\gl@bal\linesabove=\linesabove
    \s@vepicture \setbox\picturebox=\vbox{
         \kern \linesabove\baselineskip \kern 0.3\baselineskip
         \lastpicture \kern 0.3\baselineskip }%
    \dimen@=\p@cht \dimen@i=\dimen@
    \advance\dimen@i by \pagetotal
    \par \ifdim\dimen@i>\pagegoal \vfil\break \fi
    \dimen@ii=\hsize
    \advance\dimen@ii by -\parindent \advance\dimen@ii by -\p@cwd
    \setbox0=\vbox to\z@{\kern-\baselineskip \unvbox\picturebox \vss }}
\def\wr@paround{\Caption={}\count255=1
    \loop \ifnum \linesabove >0
         \advance\linesabove by -1 \advance\count255 by 1
         \advance\dimen@ by -\baselineskip
         \expandafter\Caption \expandafter{\the\Caption \z@ \hsize }%
      \repeat
    \loop \ifdim \dimen@ >\z@
         \advance\count255 by 1 \advance\dimen@ by -\baselineskip
         \expandafter\Caption \expandafter{%
             \the\Caption \dimen@i \dimen@ii }%
      \repeat
    \edef\n@xt{\parshape=\the\count255 \the\Caption \z@ \hsize }%
    \par\noindent \n@xt \strut \vadjust{\box\picturebox }}
\let\picturedefault=\relax
\let\parsefilename=\relax
\def\redopicturebox{\let\picturedefinition=\rel@x
   \errhelp=\disabledpictures
   \errmessage{This version of TeX cannot handle pictures.  Sorry.}}
\newhelp\disabledpictures
     {You will get a blank box in place of your picture.}
%
%
%
%
%
%
%
%
%
%
\def\FRONTPAGE{\ifvoid255\else\vfill\penalty-20000\fi
   \gl@bal\pagenumber=1     \gl@bal\chapternumber=0
   \gl@bal\equanumber=0     \gl@bal\sectionnumber=0
   \gl@bal\referencecount=0 \gl@bal\figurecount=0
   \gl@bal\tablecount=0     \gl@bal\frontpagetrue
   \gl@bal\lastf@@t=0       \gl@bal\footsymbolcount=0
   \gl@bal\cn@@false }
\let\Frontpage=\FRONTPAGE
\def\papers{\papersize\headline=\paperheadline\footline=\paperfootline}
\def\papersize{\hsize=35pc \vsize=50pc \hoffset=0pc \voffset=1pc
   \advance\hoffset by\HOFFSET \advance\voffset by\VOFFSET
   \pagebottomfiller=0pc
   \skip\footins=\bigskipamount \normalspace }
\papers  
%
%
\newskip\lettertopskip       \lettertopskip=20pt plus 50pt
\newskip\letterbottomskip    \letterbottomskip=\z@ plus 100pt
\newskip\signatureskip       \signatureskip=40pt plus 3pt
\def\lettersize{\hsize=6.5in \vsize=8.5in \hoffset=0in \voffset=0.5in
   \advance\hoffset by\HOFFSET \advance\voffset by\VOFFSET
   \pagebottomfiller=\letterbottomskip
   \skip\footins=\smallskipamount \multiply\skip\footins by 3
   \singlespace }
\def\MEMO{\lettersize \headline=\letterheadline \footline={\hfil }%
   \let\rule=\memorule \FRONTPAGE \memohead }

\def\memodate{\afterassignment\MEMO \date }
\def\memit@m#1{\smallskip \hangafter=0 \hangindent=1in
    \Textindent{\caps #1}}
\def\subject{\memit@m{Subject:}}
\def\topic{\memit@m{Topic:}}
\def\from{\memit@m{From:}}
\def\to{\rel@x \ifmmode \rightarrow \else \memit@m{To:}\fi }
\def\memorule{\medskip\hrule height 1pt\bigskip}  
\def\memohead{\centerline{\fourteenrm MEMORANDUM}}
\newwrite\labelswrite
\newtoks\rw@toks
\def\letters{\lettersize
   \headline=\letterheadline \footline=\letterfootline
   \immediate\openout\labelswrite=\jobname.lab}

\let\letterhead=\rel@x
\def\addressee#1{\medskip\line{\hskip 0.75\hsize plus\z@ minus 0.25\hsize
                               \the\date \hfil }%
   \vskip \lettertopskip
   \ialign to\hsize{\strut ##\hfil\tabskip 0pt plus \hsize \crcr #1\crcr}
   \writelabel{#1}\medskip \noindent\hskip -\spaceskip \ignorespaces }
\def\rwl@begin#1\cr{\rw@toks={#1\crcr}\rel@x
   \immediate\write\labelswrite{\the\rw@toks}\futurelet\n@xt\rwl@next}
\def\rwl@next{\ifx\n@xt\rwl@end \let\n@xt=\rel@x
      \else \let\n@xt=\rwl@begin \fi \n@xt}
\let\rwl@end=\rel@x
\def\writelabel#1{\immediate\write\labelswrite{\noexpand\labelbegin}
     \rwl@begin #1\cr\rwl@end
     \immediate\write\labelswrite{\noexpand\labelend}}
\newtoks\FromAddress         \FromAddress={}
\newtoks\sendername          \sendername={}
\newbox\FromLabelBox
\newdimen\labelwidth          \labelwidth=6in
\def\makelabels{\afterassignment\Makelabels \sendername=}
\def\Makelabels{\FRONTPAGE \letterinfo={\hfil } \MakeFromBox
     \immediate\closeout\labelswrite  \input \jobname.lab\vfil\eject}
\let\labelend=\rel@x
\def\labelbegin#1\labelend{\setbox0=\vbox{\ialign{##\hfil\cr #1\crcr}}
     \MakeALabel }
\def\MakeFromBox{\gl@bal\setbox\FromLabelBox=\vbox{\Tenpoint
     \ialign{##\hfil\cr \the\sendername \the\FromAddress \crcr }}}
\def\MakeALabel{\vskip 1pt \hbox{\vrule \vbox{
        \hsize=\labelwidth \hrule\bigskip
        \leftline{\hskip 1\parindent \copy\FromLabelBox}\bigskip
        \centerline{\hfil \box0 } \bigskip \hrule
        }\vrule } \vskip 1pt plus 1fil }
\def\signed#1{\par \nobreak \bigskip \dt@pfalse \begingroup
  \everycr={\noalign{\nobreak
            \ifdt@p\vskip\signatureskip\gl@bal\dt@pfalse\fi }}%
  \tabskip=0.5\hsize plus \z@ minus 0.5\hsize
  \halign to\hsize {\strut ##\hfil\tabskip=\z@ plus 1fil minus \z@\crcr
          \noalign{\gl@bal\dt@ptrue}#1\crcr }%
  \endgroup \bigskip }
\newbox\letterb@x
\def\lettertext{\par \vskip\parskip \unvcopy\letterb@x \par }
\def\multiletter{\setbox\letterb@x=\vbox\bgroup
      \everypar{\vrule height 1\baselineskip depth 0pt width 0pt }
      \singlespace \topskip=\baselineskip }
\def\letterend{\par\egroup}
%
%
%
\newskip\frontpageskip
\newtoks\Pubnum   
\newtoks\Pubtype  \let\pubtype=\Pubtype
\newif\ifp@bblock  \p@bblocktrue
\def\PH@SR@V{\doubl@true \baselineskip=24.1pt plus 0.2pt minus 0.1pt
             \parskip= 3pt plus 2pt minus 1pt }
\def\PHYSREV{\papers\PhysRevtrue\PH@SR@V}

\def\titlepage{\FRONTPAGE\papers\ifPhysRev\PH@SR@V\fi
   \ifp@bblock\p@bblock \else\hrule height\z@ \rel@x \fi }
\def\nopubblock{\p@bblockfalse}
\def\endpage{\vfil\break}
\frontpageskip=12pt plus .5fil minus 2pt
\Pubtype={}
\Pubnum={}
\def\p@bblock{\begingroup \tabskip=\hsize minus \hsize
   \baselineskip=1.5\ht\strutbox \topspace-2\baselineskip
   \halign to\hsize{\strut ##\hfil\tabskip=0pt\crcr
       \the\Pubnum\crcr\the\date\crcr\the\pubtype\crcr}\endgroup}
\def\title#1{\vskip\frontpageskip \titlestyle{#1} \vskip\headskip }
\def\author#1{\vskip\frontpageskip\titlestyle{\twelvecp #1}\nobreak}

\def\address#1{\par\kern 5pt\titlestyle{\twelvepoint\it #1}}
\def\andaddress{\par\kern 5pt \centerline{\sl and} \address}

\def\abstract{\par\dimen@=\prevdepth \hrule height\z@ \prevdepth=\dimen@
   \vskip\frontpageskip\centerline{\fourteenrm ABSTRACT}\vskip\headskip }

%
%
%

\def\\{\rel@x \ifmmode \backslash \else {\tt\char`\\}\fi }
\def\sequentialequations{\rel@x \if\equanumber<0 \else
  \gl@bal\equanumber=-\equanumber \gl@bal\advance\equanumber by -1 \fi }
\def\journal#1&#2(#3){\begingroup \let\journal=\dummyj@urnal
    \unskip, \sl #1\unskip~\bf\ignorespaces #2\rm
    (\afterassignment\j@ur \count255=#3), \endgroup\ignorespaces }
\def\j@ur{\ifnum\count255<100 \advance\count255 by 1900 \fi
          \number\count255 }
\def\dummyj@urnal{%
    \toks@={Reference foul up: nested \journal macros}%
    \errhelp={Your forgot & or ( ) after the last \journal}%
    \errmessage{\the\toks@ }}
\def\cropen#1{\crcr\noalign{\vskip #1}}

\def\topspace{\hrule height 0pt depth 0pt \vskip}

\def\Buildrel#1\under#2{\mathrel{\mathop{#2}\limits_{#1}}}
\def\becomes#1{\mathchoice{\becomes@\scriptstyle{#1}}
   {\becomes@\scriptstyle{#1}} {\becomes@\scriptscriptstyle{#1}}
   {\becomes@\scriptscriptstyle{#1}}}
\def\becomes@#1#2{\mathrel{\setbox0=\hbox{$\m@th #1{\,#2\,}$}%
        \mathop{\hbox to \wd0 {\rightarrowfill}}\limits_{#2}}}

\let\int=\intop         
\def\lsim{\mathrel{\mathpalette\@versim<}}
\def\gsim{\mathrel{\mathpalette\@versim>}}
\def\@versim#1#2{\vcenter{\offinterlineskip
        \ialign{$\m@th#1\hfil##\hfil$\crcr#2\crcr\sim\crcr } }}
\def\big#1{{\hbox{$\left#1\vbox to 0.85\b@gheight{}\right.\n@space$}}}
\def\Big#1{{\hbox{$\left#1\vbox to 1.15\b@gheight{}\right.\n@space$}}}
\def\bigg#1{{\hbox{$\left#1\vbox to 1.45\b@gheight{}\right.\n@space$}}}
\def\Bigg#1{{\hbox{$\left#1\vbox to 1.75\b@gheight{}\right.\n@space$}}}
\def\){\mskip 2mu\nobreak }
%
%
%
\let\sec@nt=\sec
\def\sec{\rel@x\ifmmode\let\n@xt=\sec@nt\else\let\n@xt\section\fi\n@xt}
\def\obsolete#1{\message{Macro \string #1 is obsolete.}}
\def\firstsec#1{\obsolete\firstsec \section{#1}}
\def\firstsubsec#1{\obsolete\firstsubsec \subsection{#1}}
\def\thispage#1{\obsolete\thispage \gl@bal\pagenumber=#1\frontpagefalse}
\def\thischapter#1{\obsolete\thischapter \gl@bal\chapternumber=#1}
\def\splitout{\obsolete\splitout\rel@x}
\def\prop{\obsolete\prop \propto }
\def\nextequation#1{\obsolete\nextequation \gl@bal\equanumber=#1
   \ifnum\the\equanumber>0 \gl@bal\advance\equanumber by 1 \fi}
\def\BOXITEM{\afterassigment\B@XITEM\setbox0=}
\def\B@XITEM{\par\hangindent\wd0 \noindent\box0 }
%
%
%
\def\phyzzx{PHY\setbox0=\hbox{Z}\copy0 \kern-0.5\wd0 \box0 X}
        
\everyjob{\xdef\today{\monthname~\number\day, \number\year}
        \input myphyx.tex }
\message{ by V.K.}
%

%
%
%
%
\def\slacpub{\afterassignment\slacp@b\toks@}
\def\slacp@b{\edef\n@xt{\Pubnum={SLAC--PUB--\the\toks@}}\n@xt}

\expandafter\ifx\csname eightrm\endcsname\relax
    \let\eightrm=\ninerm  \fi
\def\memohead{\hrule height\z@ \kern -0.5in
    \line{\quad\fourteenrm SLAC MEMORANDUM\hfil \twelverm\the\date\quad}}
\def\memorule{\par \medskip \hrule height 0.5pt \kern 1.5pt
   \hrule height 0.5pt \medskip}
\def\SLACHEAD{\setbox0=\vtop{\baselineskip=10pt
     \ialign{\eightrm ##\hfil\cr
        \slacbin\cr
        P.~O.~Box 4349\cr
        Stanford, CA 94309\cropen{1\jot}
        \slacphone\cr }}%
   \setbox2=\hbox{\caps Stanford Linear Accelerator Center}%
   \hrule height \z@ \kern -0.5in
   \vbox to 0pt{\vss\centerline{\seventeenrm STANFORD UNIVERSITY}}
   \vbox{} \medskip
   \line{\hbox to 0.7\hsize{\hss \lower 10pt \box2 \hfill }\hfil
         \hbox to 0.25\hsize{\box0 \hfil }}\medskip }
\let\letterhead=\SLACHEAD
\FromAddress={\crcr \slacbin \cr
    P.\ O.\ Box 4349\cr Stanford, California 94309\cr }
   \def\slacbin{%
      SLAC%
      \expandafter\ifx\csname binno\endcsname\relax%
         \expandafter\ifx\csname MailStop\endcsname\relax%
         \else%
            , Mail Stop \MailStop%
         \fi%
      \else%
         , Mail Stop \binno%
      \fi%
   }
\def\slacphone{(415) 926--\slacext}
\def\slacext{3300}
\VOFFSET=33pt
\papersize
%
%
\newwrite\figscalewrite
\newif\iffigscaleopen
\newif\ifgrayscale
\newif\ifreadyfile
\def\picturedefault{\grayscalefalse \readyfilefalse
    \gdef\ready{\readyfiletrue}\gdef\gray{\ready\grayscaletrue}}
\def\parsefilename{\ifreadyfile \else
    \iffigscaleopen \else \gl@bal\figscaleopentrue
       \immediate\openout\figscalewrite=\jobname.scalecon \fi
    \toks0={ }\immediate\write\figscalewrite{%
       \the\p@cwd \the\toks0 \the\p@cht \the\toks0 \the\picfilename }%
    \expandafter\p@rse \the\picfilename..\endp@rse \fi }
\def\p@rse#1.#2.#3\endp@rse{%
   \if*#3*\dop@rse #1.1..\else \if.#3\dop@rse #1.1..\else
                                \dop@rse #1.#3\fi \fi
   \expandafter\picfilename\expandafter{\n@xt}}
\def\dop@rse#1.#2..{\count255=#2 \ifnum\count255<1 \count255=1 \fi
   \ifnum\count255<10  \edef\n@xt{#1.PICT00\the\count255}\else
   \ifnum\count255<100 \edef\n@xt{#1.PICT0\the\count255}\else
                       \edef\n@xt{#1.PICT\the\count255}\fi\fi }
\def\redopicturebox{\edef\picturedefinition{\ifgrayscale
     \special{insert(\the\picfilename)}\else
     \special{mergeug(\the\picfilename)}\fi }}
%
%
\let\us=\undertext
\let\rule=\memorule
\let\eqnalign=\eqname
\def\boxit#1{\vbox{\hrule\hbox{\vrule\kern3pt
\vbox{\kern3pt#1\kern3pt}\kern3pt\vrule}\hrule}}

%
%
   \message{V 1.18 mods and bug fixes by M.Weinstein}
   \def\unlock{\catcode`@=11}

   \def\lock{\catcode`@=12}

   \unlock
%
%
   \def\PRrefmark#1{~[#1]}
   \def\refitem#1{\ifPhysRev\r@fitem{[#1]}\else\r@fitem{#1.}\fi}
   \def\generatefootsymbol{%
      \ifcase\footsymbolcount%
          \fd@f 13F \or \fd@f 279 \or \fd@f 27A %
            \or \fd@f 278 \or \fd@f 27B %
      \else
         \ifnum\footsymbolcount <0 %
            \xdef\footsymbol{\number-\footsymbolcount}
         \else %
            \fd@f 203
               {\loop \ifnum\footsymbolcount >5
                  \fd@f{203 \footsymbol } %
                  \advance\footsymbolcount by -1%
                \repeat %
               }%
         \fi%
      \fi%
   }
   \def\OldPhysRevRefmark{\let\PRrefmark=\attach}
   \def\OldPRRefitem#1{\r@fitem{#1.}}
   \def\OldPhysRevRefitem{\let\refitem=\OldPRRefitem}
   \def\NPrefs{\let\refmark=\NPrefmark \let\refitem=\NPrefitem}
%
    \newif\iffileexists              \fileexistsfalse
    \newif\ifforwardrefson           \forwardrefsontrue
    \newif\ifamiga                   \amigafalse
    \newif\iflinkedinput             \linkedinputtrue
    \newif\iflinkopen                \linkopenfalse
    \newif\ifcsnameopen              \csnameopenfalse
    \newif\ifdummypictures           \dummypicturesfalse
    \newif\ifcontentson              \contentsonfalse
    \newif\ifcontentsopen            \contentsopenfalse
    \newif\ifmakename                \makenamefalse
    \newif\ifverbdone
    \newif\ifusechapterlabel         \usechapterlabelfalse
    \newif\ifstartofchapter          \startofchapterfalse
    \newif\iftableofplates           \tableofplatesfalse
    \newif\ifplatesopen              \platesopenfalse
    \newif\iftableoftables           \tableoftablesfalse
    \newif\iftableoftablesopen       \tableoftablesopenfalse
    \newif\ifwarncsname              \warncsnamefalse
%
    \newwrite\linkwrite
    \newwrite\csnamewrite
    \newwrite\contentswrite
    \newwrite\plateswrite
    \newwrite\tableoftableswrite
    \newread\testifexists
    \newread\verbinfile

    \newtoks\jobdir                  \jobdir={}
    \newtoks\tempnametoks            \tempnametoks={}
    \newtoks\oldheadline             \oldheadline={}
    \newtoks\oldfootline             \oldfootline={}
    \newtoks\subsectstyle            \subsectstyle={\Number}
    \newtoks\subsubsectstyle         \subsubsectstyle={\Number}
    \newtoks\runningheadlines        \runningheadlines={\relax}
    \newtoks\chapterformat           \chapterformat={\titlestyle}
    \newtoks\sectionformat           \sectionformat={\relax}
    \newtoks\subsectionformat        \subsectionformat={\relax}
    \newtoks\subsubsectionformat     \subsubsectionformat={\relax}
    \newtoks\chapterfontstyle        \chapterfontstyle={\bf}
    \newtoks\sectionfontstyle        \sectionfontstyle={\rm}
    \newtoks\subsectionfontstyle     \subsectionfontstyle={\rm}
    \newtoks\sectionfontstyleb       \sectionfontstyleb={\caps}
    \newtoks\subsubsectionfontstyle  \subsubsectionfontstyle={\rm}

    \newcount\subsectnumber           \subsectnumber=0
    \newcount\subsubsectnumber        \subsubsectnumber=0


   \newdimen\pictureindent           \pictureindent=15pt
   \newdimen\str
   \newdimen\squareht
   \newdimen\squarewd
   \newskip\doublecolskip
   \newskip\tableoftablesskip        \tableoftablesskip=\baselineskip


   \newbox\squarebox


   \newskip\sectionindent            \sectionindent=0pt
   \newskip\subsectionindent         \subsectionindent=0pt
  \def\thechapterhead{\relax}
  \def\thesectionhead{\relax}
  \def\thesubsecthead{\relax}
  \def\thesubsubsecthead{\relax}


   \def\GetIfExists #1 {
       \immediate\openin\testifexists=#1
       \ifeof\testifexists
           \immediate\closein\testifexists
       \else
         \immediate\closein\testifexists
         \input #1
       \fi
   }


   \def\stripbackslash#1#2*{\def\strippedname{#2}}

   \def\ifundefined#1{\expandafter\ifx\csname#1\endcsname\relax}

   \def\val#1{%
      \expandafter\stripbackslash\string#1*%
      \ifundefined{\strippedname}%
      \message{Warning! The control sequence \noexpand#1 is not defined.} ? %
      \else\csname\strippedname\endcsname\fi%
   }
%
%
   \def\CheckForOverWrite#1{%
      \expandafter\stripbackslash\string#1*%
      \ifundefined{\strippedname}%
      \else%
         \ifwarncsname
            \message{Warning! The control sequence \noexpand#1 is being
          overwritten.}%
          \else
          \fi
      \fi%
   }

   \def\FootNoteFonts{\Tenpoint}

   \def\Vfootnote#1{%
      \insert\footins%
      \bgroup%
         \interlinepenalty=\interfootnotelinepenalty%
         \floatingpenalty=20000%
         \singl@true\doubl@false%
         \FootNoteFonts%
         \splittopskip=\ht\strutbox%
         \boxmaxdepth=\dp\strutbox%
         \leftskip=\footindent%
         \rightskip=\z@skip%
         \parindent=0.5%
         \footindent%
         \parfillskip=0pt plus 1fil%
         \spaceskip=\z@skip%
         \xspaceskip=\z@skip%
         \footnotespecial%
         \Textindent{#1}%
         \footstrut%
         \futurelet\next\fo@t%
   }

   \def\csnamech@ck{%
       \ifcsnameopen%
       \else%
           \global\csnameopentrue%
           \immediate\openout\csnamewrite=\the\jobdir\jobname.csnames%
           \immediate\write\csnamewrite{\unlock}%
       \fi%
   }

   \def\linksch@ck{%
          \iflinkopen%
          \else%
              \global\linkopentrue%
              \immediate\openout\linkwrite=\the\jobdir\jobname.links%
          \fi%
   }

   \def\c@ntentscheck{%
       \ifcontentsopen%
       \else%
           \global\contentsopentrue%
           \immediate\openout\contentswrite=\the\jobdir\jobname.contents%
           \immediate\write\contentswrite{%
                \noexpand\titlestyle{Table of Contents}%
           }%
           \immediate\write\contentswrite{\noexpand\bigskip}%
       \fi%
   }

   \def\t@bleofplatescheck{%
       \ifplatesopen%
       \else%
           \global\platesopentrue%
           \immediate\openout\plateswrite=\the\jobdir\jobname.plates%
           \immediate\write\plateswrite{%
                \noexpand\titlestyle{Illustrations}%
           }%
           \immediate\write\plateswrite{%
              \unlock%
           }%
           \immediate\write\plateswrite{\noexpand\bigskip}%
       \fi%
   }

   \def\t@bleoftablescheck{%
       \iftableoftablesopen%
       \else%
           \global\tableoftablesopentrue%
          \immediate\openout\tableoftableswrite=\the\jobdir\jobname.tables%
           \immediate\write\tableoftableswrite{%
                \noexpand\titlestyle{Tables}%
           }%
           \immediate\write\tableoftableswrite{%
              \unlock%
           }%
           \immediate\write\tableoftableswrite{\noexpand\bigskip}%
       \fi%
   }


   \def\linkinput#1 {\input #1
       \iflinkedinput \relax \else \global\linkedinputtrue \fi
       \linksch@ck
       \immediate\write\linkwrite{#1}
   }


   \def\fil@#1 {%
       \ifdummypictures%
          \fileexistsfalse%
          \picfilename={}%
       \else%
          \fileexiststrue%
          \picfilename={#1}%
       \fi%
       \iflinkedinput%
          \iflinkopen \relax%
          \else%
            \global\linkopentrue%
            \immediate\openout\linkwrite=\the\jobdir\jobname.links%
          \fi%
          \immediate\write\linkwrite{#1}%
       \fi%
   }
   \def\Picture#1{%
      \gl@bal\advance\figurecount by 1%
      \CheckForOverWrite#1%
      \xdef#1{\the\figurecount}\pl@t%
      \selfcaptionedtrue%
   }

   \def\s@vepicture{%
       \iffileexists \parsefilename \redopicturebox \fi%
       \ifdim\captionwidth>\z@ \else \captionwidth=\p@cwd \fi%
       \xdef\lastpicture{%
          \iffileexists%
             \setbox0=\hbox{\raise\the\yshift \vbox{%
                \moveright\the\xshift\hbox{\picturedefinition}}%
             }%
          \else%
             \setbox0=\hbox{}%
          \fi
          \ht0=\the\p@cht \wd0=\the\p@cwd \dp0=\the\p@cdp%
          \vbox{\hsize=\the\captionwidth \line{\hss\box0 \hss }%
          \ifcaptioned%
             \vskip\the\captionskip \noexpand\Tenpoint%
             \ifselfcaptioned%
                Figure~\the\figurecount.\enspace%
             \fi%
             \the\Caption%
          \fi }%
       }%
       \iftableofplates%
          \ifplatesopen%
          \else%
             \t@bleofplatescheck%
          \fi%
          \ifselfcaptioned%
             \immediate\write\plateswrite{%
                \noexpand\platetext{%
                \noexpand\item{\rm \the\figurecount .}%
                \the\Caption}{\the\pageno}%
             }%
          \else%
             \immediate\write\plateswrite{%
                \noexpand\platetext{\the\Caption}{\the\pageno}%
             }%
          \fi%
       \fi%
       \endgroup%
   }

   \def\platesout{%
      \ifplatesopen
         \immediate\closeout\plateswrite%
         \global\platesopenfalse%
      \fi%
      \input \jobname.plates%
      \lock%
   }

   \def\platetext#1#2{%
       \hbox to \hsize{\vbox{\hsize=.9\hsize #1}\hfill#2}%
       \vskip \tableoftablesskip \vskip\parskip%
   }


   \def\pres@tpicture{%
       \gl@bal\linesabove=\linesabove
       \s@vepicture
       \setbox\picturebox=\vbox{
       \kern \linesabove\baselineskip \kern 0.3\baselineskip
       \lastpicture \kern 0.3\baselineskip }%
       \dimen@=\p@cht \dimen@i=\dimen@
       \advance\dimen@i by \pagetotal
       \par \ifdim\dimen@i>\pagegoal \vfil\break \fi
       \dimen@ii=\hsize
       \advance\dimen@ii by -\pictureindent \advance\dimen@ii by -\p@cwd
       \setbox0=\vbox to\z@{\kern-\baselineskip \unvbox\picturebox \vss }
   }

   \def\subspaces@t#1:#2;{%
      \baselineskip = \normalbaselineskip%
      \multiply\baselineskip by #1 \divide\baselineskip by #2%
      \lineskip = \normallineskip%
      \multiply\lineskip by #1 \divide\lineskip by #2%
      \lineskiplimit = \normallineskiplimit%
      \multiply\lineskiplimit by #1 \divide\lineskiplimit by #2%
      \parskip = \normalparskip%
      \multiply\parskip by #1 \divide\parskip by #2%
      \abovedisplayskip = \normaldisplayskip%
      \multiply\abovedisplayskip by #1 \divide\abovedisplayskip by #2%
      \belowdisplayskip = \abovedisplayskip%
      \abovedisplayshortskip = \normaldispshortskip%
      \multiply\abovedisplayshortskip by #1%
        \divide\abovedisplayshortskip by #2%
      \belowdisplayshortskip = \abovedisplayshortskip%
      \advance\belowdisplayshortskip by \belowdisplayskip%
      \divide\belowdisplayshortskip by 2%
      \smallskipamount = \skipregister%
      \multiply\smallskipamount by #1 \divide\smallskipamount by #2%
      \medskipamount = \smallskipamount \multiply\medskipamount by 2%
      \bigskipamount = \smallskipamount \multiply\bigskipamount by 4%
   }


   \def\makename#1{
       \global\makenametrue
       \global\tempnametoks={#1}
   }

   \def\nomakename#1{\relax}


   \def\savename#1{%
      \CheckForOverWrite{#1}%
      \csnamech@ck%
      \immediate\write\csnamewrite{\def\the\tempnametoks{#1}}%
   }

   \def\FootNoteFonts{\Tenpoint}

   \def\Vfootnote#1{%
      \insert\footins%
      \bgroup%
         \interlinepenalty=\interfootnotelinepenalty%
         \floatingpenalty=20000%
         \singl@true\doubl@false%
         \FootNoteFonts%
         \splittopskip=\ht\strutbox%
         \boxmaxdepth=\dp\strutbox%
         \leftskip=\footindent%
         \rightskip=\z@skip%
         \parindent=0.5%
         \footindent%
         \parfillskip=0pt plus 1fil%
         \spaceskip=\z@skip%
         \xspaceskip=\z@skip%
         \footnotespecial%
         \Textindent{#1}%
         \footstrut%
         \futurelet\next\fo@t%
   }
%

   \def\eqname#1{%
      \CheckForOverWrite{#1}%
      \rel@x{\pr@tect%
      \csnamech@ck%
      \ifnum\equanumber<0%
          \xdef#1{{\noexpand\f@m0(\number-\equanumber)}}%
          \immediate\write\csnamewrite{%
            \def\noexpand#1{\noexpand\f@m0 (\number-\equanumber)}}%
          \gl@bal\advance\equanumber by -1%
      \else%
          \gl@bal\advance\equanumber by 1%
          \ifusechapterlabel%
            \xdef#1{{\noexpand\f@m0(\ifcn@@ \chapterlabel.\fi%
               \number\equanumber)}%
            }%
          \else%
             \xdef#1{{\noexpand\f@m0(\ifcn@@%
                 {\the\chapterstyle{\the\chapternumber}}.\fi%
                 \number\equanumber)}}%
          \fi%
          \ifcn@@%
             \ifusechapterlabel
                \immediate\write\csnamewrite{\def\noexpand#1{(%
                  {\chapterlabel}.%
                  \number\equanumber)}%
                }%
             \else
                \immediate\write\csnamewrite{\def\noexpand#1{(%
                  {\the\chapterstyle{\the\chapternumber}}.%
                  \number\equanumber)}%
                }%
             \fi%
          \else%
              \immediate\write\csnamewrite{\def\noexpand#1{(%
                  \number\equanumber)}}%
          \fi%
      \fi}%
      #1%
   }

   \def\eq{\eqname\?}

   \def\eqn{\eqno\eqname}

   \let\eqnalign=\eqname


   \def\APPENDIX#1#2{%
       \global\usechapterlabeltrue%
       \par\penalty-300\vskip\chapterskip%
       \spacecheck\chapterminspace%
       \chapterreset%
       \xdef\chapterlabel{#1}%
       \titlestyle{APPENDIX #2}%
       \nobreak\vskip\headskip \penalty 30000%
       \wlog{\string\Appendix~\chapterlabel}%
   }

   \def\REFNUM#1{%
      \CheckForOverWrite{#1} %
      \rel@x\gl@bal\advance\referencecount by 1%
      \xdef#1{\the\referencecount}%
      \csnamech@ck%
      \immediate\write\csnamewrite{\def\noexpand#1{\the\referencecount}}%
   }

   %

   \def\FIGNUM#1{
      \CheckForOverWrite{#1}%
      \rel@x\gl@bal\advance\figurecount by 1%
      \xdef#1{\the\figurecount}%
      \csnamech@ck%
      \immediate\write\csnamewrite{\def\noexpand#1{\the\figurecount}}%
   }


   \def\TABNUM#1{%
      \CheckForOverWrite{#1}%
      \rel@x \gl@bal\advance\tablecount by 1%
      \xdef#1{\the\tablecount}%
      \csnamech@ck%
      \immediate\write\csnamewrite{\def\noexpand#1{\the\tablecount}}%
   }


   \def\tableoftableson{%
      \global\tableoftablestrue%

      \gdef\TABLE##1##2{%
         \t@bleoftablescheck%
         \TABNUM ##1%
         \immediate\write\tableoftableswrite{%
            \noexpand\tableoftablestext{%
            \noexpand\item{\rm \the\tablecount .}%
                ##2}{\the\pageno}%
             }%
      }

      \gdef\Table##1{\TABLE\?{##1}Table~\?}
   }

   \def\tableoftablestext#1#2{%
       \hbox to \hsize{\vbox{\hsize=.9\hsize #1}\hfill#2}%
       \vskip \tableoftablesskip%
   }

   \def\tableoftablesout{%
      \iftableoftablesopen
         \immediate\closeout\tableoftableswrite%
         \global\tableoftablesopenfalse%
      \fi%
      \input \jobname.tables%
      \lock%
   }

%
%
%
%
%
%

   \def\contentsoff{\contentsonfalse}

   \def\f@m#1{\f@ntkey=#1\fam=\f@ntkey\the\textfont\f@ntkey\rel@x}
   \def\em@{\rel@x%
      \ifnum\f@ntkey=0\it%
      \else%
         \ifnum\f@ntkey=\bffam\it%
         \else\rm  %
         \fi%
      \fi%
   }

   \def\fontsoff{%
      \def\mit{\relax}%
      \let\oldstyle=\mit%
      \def\cal{\relax}%
      \def\it{\relax}%
      \def\sl{\relax}%
      \def\bf{\relax}%
      \def\tt{\relax}%
      \def\caps{\relax}%
      \let\cp=\caps%
   }


   \def\fontson{%
      \def\rm{\n@expand\f@m0}%
      \def\mit{\n@expand\f@m1}%
      \let\oldstyle=\mit%
      \def\cal{\n@expand\f@m2}%
      \def\it{\n@expand\f@m\itfam}%
      \def\sl{\n@expand\f@m\slfam}%
      \def\bf{\n@expand\f@m\bffam}%
      \def\tt{\n@expand\f@m\ttfam}%
      \def\caps{\n@expand\f@m\cpfam}%
      \let\cp=\caps%
   }

   \fontson
%


   \def\@alpha#1{\count255='140 \advance\count255 by #1\char\count255}
   \def\alphabetic{\@alpha}
   \def\@Alpha#1{\count255='100 \advance\count255 by #1\char\count255}
   \def\Alphabetic{\@Alpha}
   \def\@Roman#1{\uppercase\expandafter{\romannumeral #1}}
   \def\Roman{\@Roman}
   \def\@roman#1{\romannumeral #1}
   \def\roman{\@roman}
   \def\@number#1{\number #1}
   \def\Number{\@number}

   \def\leaderfill{\leaders\hbox to 1em{\hss.\hss}\hfill}

   \def\chapterinfo#1{%
      \line{%
         \ifcn@@%
            \hbox to \itemsize{\hfil\chapterlabel .\quad\ }%
         \fi%
         \noexpand{#1}\leaderfill\the\pagenumber%
      }%
   }

   \def\sectioninfo#1{%
      \line{%
         \ifcn@@%
            \hbox to 2\itemsize{\hfil\sectlabel \quad}%
          \else%
            \hbox to \itemsize{\hfil\quad}%
          \fi%
          \ \noexpand{#1}%
          \leaderfill \the\pagenumber%
      }%
   }

   \def\subsectioninfo#1{%
      \line{%
         \ifcn@@%
            \hbox to 3\itemsize{\hfil \quad\subsectlabel\quad}%
         \else%
            \hbox to 2\itemsize{\hfil\quad}%
         \fi%
          \ \noexpand{#1}%
          \leaderfill \the\pagenumber%
      }%
   }

   \def\subsubsecinfo#1{%
      \line{%
         \ifcn@@%
            \hbox to 4\itemsize{\hfil\subsubsectlabel\quad}%
         \else%
            \hbox to 3\itemsize{\hfil\quad}%
         \fi%
         \ \noexpand{#1}\leaderfill \the\pagenumber%
      }%
   }

   \def\CONTENTS#1;#2{
       {\let\makename=\nomakename
        \if#1C
            \immediate\write\contentswrite{\chapterinfo{#2}}%
        \else\if#1S
                \immediate\write\contentswrite{\sectioninfo{#2}}%
             \else\if#1s
                     \immediate\write\contentswrite{\subsectioninfo{#2}}%
                  \else\if#1x
                          \immediate\write\contentswrite{%
                              \subsubsecinfo{#2}}%
                       \fi
                  \fi
             \fi
        \fi
       }
   }

   \def\chapterreset{\gl@bal\advance\chapternumber by 1%
       \ifnum\equanumber<0 \else\gl@bal\equanumber=0 \fi%
       \gl@bal\sectionnumber=0  \gl@bal\let\sectlabel=\rel@x%
       \gl@bal\subsectnumber=0   \gl@bal\let\subsectlabel=\rel@x%
       \gl@bal\subsubsectnumber=0 \gl@bal\let\subsubsectlabel=\rel@x%
       \ifcn@%
           \gl@bal\cn@@true {\pr@tect\xdef\chapterlabel{%
           {\the\chapterstyle{\the\chapternumber}}}}%
       \else%
           \gl@bal\cn@@false \gdef\chapterlabel{\rel@x}%
       \fi%
       \gl@bal\startofchaptertrue%
   }

   \def\chapter#1{\par \penalty-300 \vskip\chapterskip%
       \spacecheck\chapterminspace%
       \gdef\thechapterhead{#1}%
       \gdef\thesectionhead{\relax}%
       \gdef\thesubsecthead{\relax}%
       \gdef\thesubsubsecthead{\relax}%
       \chapterreset \the\chapterformat{\the\chapterfontstyle%
          \ifcn@@\chapterlabel.~~\fi #1}%
       \nobreak\vskip\headskip \penalty 30000%
       {\pr@tect\wlog{\string\chapter\space \chapterlabel}}%
       \ifmakename%
           \csnamech@ck
           \ifcn@@%
              \immediate\write\csnamewrite{\def\the\tempnametoks{%
                 {\the\chapterstyle{\the\chapternumber}}}%
              }%
            \fi%
            \global\makenamefalse%
       \fi%
       \ifcontentson%
          \c@ntentscheck%
          \CONTENTS{C};{#1}%
       \fi%
       }%

   \def\section#1{\par \ifnum\lastpenalty=30000\else%
       \penalty-200\vskip\sectionskip \spacecheck\sectionminspace\fi%
       \gl@bal\advance\sectionnumber by 1%
       \gl@bal\subsectnumber=0%
       \gl@bal\let\subsectlabel=\rel@x%
       \gl@bal\subsubsectnumber=0%
       \gl@bal\let\subsubsectlabel=\rel@x%
       \gdef\thesectionhead{#1}%
       \gdef\thesubsecthead{\relax}%
       \gdef\thesubsubsecthead{\relax}%
       {\pr@tect\xdef\sectlabel{\ifcn@@%
          {\the\chapterstyle{\the\chapternumber}}.%
          {\the\sectionstyle{\the\sectionnumber}}\fi}%
       \wlog{\string\section\space \sectlabel}}%
       \the\sectionformat{\noindent\the\sectionfontstyle%
            {\ifcn@@\unskip\hskip\sectionindent\sectlabel~~\fi%
                \the\sectionfontstyleb#1}}%
       \par%
       \nobreak\vskip\headskip \penalty 30000%
       \ifmakename%
           \csnamech@ck%
           \ifcn@@%
              \immediate\write\csnamewrite{\def\the\tempnametoks{%
                 {\the\chapterstyle{\the\chapternumber}.%
                  \the\sectionstyle{\the\sectionnumber}}}
              }%
            \fi%
            \global\makenamefalse%
       \fi%
       \ifcontentson%
          \c@ntentscheck%
          \CONTENTS{S};{#1}%
       \fi%
   }

   \def\subsection#1{\par \ifnum\lastpenalty=30000\else%
       \penalty-200\vskip\sectionskip \spacecheck\sectionminspace\fi%
       \gl@bal\advance\subsectnumber by 1%
       \gl@bal\subsubsectnumber=0%
       \gl@bal\let\subsubsectlabel=\rel@x%
       \gdef\thesubsecthead{#1}%
       \gdef\thesubsubsecthead{\relax}%
       {\pr@tect\xdef\subsectlabel{\the\subsectionfontstyle%
           \ifcn@@{\the\chapterstyle{\the\chapternumber}}.%
           {\the\sectionstyle{\the\sectionnumber}}.%
           {\the\subsectstyle{\the\subsectnumber}}\fi}%
           \wlog{\string\section\space \subsectlabel}%
       }%
       \the\subsectionformat{\noindent\the\subsectionfontstyle%
         {\ifcn@@\unskip\hskip\subsectionindent%
          \subsectlabel~~\fi#1}}%
       \par%
       \nobreak\vskip\headskip \penalty 30000%
       \ifmakename%
           \csnamech@ck%
           \ifcn@@%
              \immediate\write\csnamewrite{\def\the\tempnametoks{%
                 {\the\chapterstyle{\the\chapternumber}}.%
                 {\the\sectionstyle{\the\sectionnumber}}.%
                 {\the\subsectstyle{\the\subsectnumber}}}%
              }%
            \fi%
            \global\makenamefalse%
       \fi%
       \ifcontentson%
          \c@ntentscheck%
          \CONTENTS{s};{#1}%
       \fi%
   }

   \def\subsubsection#1{\par \ifnum\lastpenalty=30000\else%
       \penalty-200\vskip\sectionskip \spacecheck\sectionminspace\fi%
       \gl@bal\advance\subsubsectnumber by 1%
       \gdef\thesubsubsecthead{#1}%
       {\pr@tect\xdef\subsubsectlabel{\the\subsubsectionfontstyle\ifcn@@%
           {\the\chapterstyle{\the\chapternumber}}.%
           {\the\sectionstyle{\the\sectionnumber}}.%
           {\the\subsectstyle{\the\subsectnumber}}.%
           {\the\subsubsectstyle{\the\subsubsectnumber}}\fi}%
           \wlog{\string\section\space \subsubsectlabel}%
       }%
       \the\subsubsectionformat{\the\subsubsectionfontstyle%
          \noindent{\ifcn@@\unskip\hskip\subsectionindent%
            \subsubsectlabel~~\fi#1}}%
       \par%
       \nobreak\vskip\headskip \penalty 30000%
       \ifmakename%
           \csnamech@ck%
           \ifcn@@%
              \immediate\write\csnamewrite{\def\the\tempnametoks{%
                {\the\chapterstyle{\the\chapternumber}.%
                 \the\sectionstyle{\the\sectionnumber}.%
                 \the\subsectionstyle{\the\subsectnumber}.%
                 \the\subsubsectstyle{\the\subsubsectnumber}}}%
              }%
            \fi%
            \global\makenamefalse%
       \fi%
       \ifcontentson%
          \c@ntentscheck%
          \CONTENTS{x};{#1}%
       \fi%
   }%

   \def\contentsinput{%
       \ifcontentson%
           \contentsopenfalse%
           \immediate\closeout\contentswrite%
           \global\oldheadline=\headline%
           \global\headline={\hfill}%
           \global\oldfootline=\footline%
           \global\footline={\hfill}%
           \fontsoff \unlock%
           \input \the\jobdir\jobname.contents%
           \fontson%
           \lock%
           \endpage%
           \global\headline=\oldheadline%
           \global\footline=\oldfootline%
       \else%
           \relax%
       \fi%
   }


       \def\phyzzxfootline{
           \footline={\ifletterstyle\the\letterfootline%
               \else\the\paperfootline\fi}%
       }

%

   {\obeyspaces}

   \def\verbfile#1{
       {\catcode`\\=12\catcode`\{=12
       \catcode`\}=12\catcode`\$=12\catcode`\&=12
       \catcode`\#=12\catcode`\%=12\catcode`\~=12
       \catcode`\_=12\catcode`\^=12\obeyspaces\obeylines\tt
       \verbdonetrue\openin\verbinfile=#1
       \loop\read\verbinfile to \inline
           \ifeof\verbinfile
               \verbdonefalse
           \else
              \leftline{\inline}
           \fi
       \ifverbdone\repeat
       \closein\verbinfile}
   }

   \def\boxit#1{\vbox{\hrule\hbox{\vrule\kern3pt%
       \vbox{\kern3pt#1\kern3pt}\kern3pt\vrule}\hrule}%
   }

   \def\square{%
      \setbox\squarebox=\boxit{\hbox{\phantom{x}}}
      \squareht = 1\ht\squarebox
      \squarewd = 1\wd\squarebox
      \vbox to 0pt{
          \offinterlineskip \kern -.9\squareht
          \hbox{\copy\squarebox \vrule width .2\squarewd height .8\squareht
              depth 0pt \hfill
          }
          \hbox{\kern .2\squarewd\vbox{%
            \hrule height .2\squarewd width \squarewd}
          }
          \vss
      }
   }

   \def\fboxit#1#2{
       \vbox{\hrule height #1
           \hbox{\vrule width #1
               \kern3pt \vbox{\kern3pt#2\kern3pt}\kern3pt \vrule width #1
           }
           \hrule height #1
       }
   }

   \let\eqnameold=\eqname

   \def\draft{\def\eqname##1{\eqnameold##1:{\tt\string##1}}
      \let\eqnalign = \eqname
   }
%
%
   \def\runningrightheadline{%
       \hfill%
       \tenit%
       \ifstartofchapter%
          \global\startofchapterfalse%
       \else%
          \ifcn@@ \the\chapternumber.\the\sectionnumber\quad\fi%
              {\fontsoff\thesectionhead}%
       \fi%
       \qquad\twelverm\folio%
   }

   \def\runningleftheadline{%
      \twelverm\folio\qquad%
      \tenit%
      \ifstartofchapter%
          \global\startofchapterfalse%
      \else%
         \ifcn@@%
             Chapter \the\chapternumber \quad%
         \fi%
         {\fontsoff\thechapterhead}%
         \hfill%
      \fi%
   }

   \runningheadlines={%
      \ifodd\pageno%
         \runningrightheadline%
      \else%
         \runningleftheadline%
      \fi
   }

%
%
%
%
%

   \font\dfont=cmr10 scaled \magstep5


   \newbox\cstrutbox
   \newbox\dlbox
   \newbox\vsk

   \setbox\cstrutbox=\hbox{\vrule height10.5pt depth3.5pt width\z@}

   \def\cstrut{\relax\ifmmode\copy\cstrutbox\else\unhcopy\cstrutbox\fi}

   \def\dl #1{\noindent\strut
       \setbox\dlbox=\hbox{\dfont #1\kern 2pt}%
       \setbox\vsk=\hbox{(}%
       \hangindent=1.1\wd\dlbox
       \hangafter=-2
       \strut\hbox to 0pt{\hss\vbox to 0pt{%
         \vskip-.75\ht\vsk\box\dlbox\vss}}%
       \noindent
   }

%
%

   \newdimen\fullhsize

   \fullhsize=6.5in
   \def\fullline{\hbox to\fullhsize}
   \let\l@r=L

   \newbox\leftcolumn
   \newbox\midcolumn

   \def\twocols{ \hsize = 3.1in
%
%
%
%

      \doublecolskip=.3333em plus .3333em minus .1em
      \global\spaceskip=\doublecolskip%
      \global\hyphenpenalty=0

      \singlespace

      \gdef\makeheadline{
          \vbox to 0pt{ \skip@=\topskip
          \advance\skip@ by -12pt \advance\skip@ by -2\normalbaselineskip
          \vskip\skip@
          \fullline{\vbox to 12pt{}\the\headline} \vss}\nointerlineskip
      }

      \def\makefootline{\baselineskip = 1.5\normalbaselineskip
           \fullline{\the\footline}
      }

      \output={
          \if L\l@r
             \global\setbox\leftcolumn=\columnbox \global\let\l@r=R
          \else
              \doubleformat \global\let\l@r=L
          \fi
          \ifnum\outputpenalty>-20000 \else\dosupereject\fi
      }

      \def\doubleformat{
          \shipout\vbox{
             \makeheadline
             \fullline{\box\leftcolumn\hfil\columnbox}
             \makefootline
          }
          \advancepageno
      }

      \def\columnbox{\leftline{\pagebody}}

      \outer\def\twobye{
          \par\vfill\supereject\if R\l@r \null\vfill\eject\fi\end
      }
   }

   \def\threecols{
       \hsize = 2.0in \tenpoint

      \doublecolskip=.3333em plus .3333em minus .1em
      \global\spaceskip=\doublecolskip%
      \global\hyphenpenalty=0

       \singlespace

       \def\makeheadline{\vbox to 0pt{ \skip@=\topskip
           \advance\skip@ by -12pt \advance\skip@ by -2\normalbaselineskip
           \vskip\skip@ \fullline{\vbox to 12pt{}\the\headline} \vss
           }\nointerlineskip
       }
       \def\makefootline{\baselineskip = 1.5\normalbaselineskip
                 \fullline{\the\footline}
       }

       \output={
          \if L\l@r
             \global\setbox\leftcolumn=\columnbox \global\let\l@r=M
          \else \if M\l@r
                   \global\setbox\midcolumn=\columnbox
                   \global\let\l@r=R
                \else \tripleformat \global\let\l@r=L
                \fi
          \fi
          \ifnum\outputpenalty>-20000 \else\dosupereject\fi
       }

       \def\tripleformat{
           \shipout\vbox{
               \makeheadline
               \fullline{\box\leftcolumn\hfil\box\midcolumn\hfil\columnbox}
               \makefootline
           }
           \advancepageno
       }

       \def\columnbox{\leftline{\pagebody}}

       \outer\def\threebye{
           \par\vfill\supereject
           \if R\l@r \null\vfill\eject\fi
           \end
       }
   }


%
%
%

   \everyjob{%
      \xdef\today{\monthname~\number\day, \number\year}
      
       \immediate\openin\testifexists=m
       \ifeof\testifexists
           \immediate\closein\testifexists
       \else
         \immediate\closein\testifexists
         \input m
       \fi
   yphyx.tex
      \ifforwardrefson%
         
       \immediate\openin\testifexists=\the
       \ifeof\testifexists
           \immediate\closein\testifexists
       \else
         \immediate\closein\testifexists
         \input \the
       \fi
   \jobdir\jobname.csnames
      \fi%
   }

\contentsoff
\lock

\catcode`\@=12 

\PHYSREV
 \newtoks\slashfraction
 \slashfraction={.13}
 \def\slash#1{\setbox0\hbox{$ #1 $}
 \setbox0\hbox to \the\slashfraction\wd0{\hss \box0}/\box0 }

 \def\str{\penalty-10000\hfilneg\ }  
 \def\nostr{\hfill\penalty-10000\ }  

\def\GeV{{\rm GeV}}
\def\MeV{{\rm MeV}}
\def\d{{\delta}}
\def\MS{{\rm \overline{MS}}}
\def\tH{{\rm 't H}}
\def\simlt
    {\hbox{\raise0.1ex\hbox{$<$}\kern-0.8em\lower0.6ex\hbox{$\sim$}}}
\def\simgt
    {\hbox{\raise0.5ex\hbox{$>$}\kern-0.8em\lower0.6ex\hbox{$\sim$}}}
\overfullrule 0pt


\Frontpage

\vbox to 8.499in{
 \vbox{\hsize=6.4in\tenrm\baselineskip 12.1pt plus .02in minus .01in
  \rightline{\hbox to 1.5in{\tenrm  SLAC--PUB--5938\hfil }}
  \rightline{\hbox to 1.5in{\tenrm October 1992\hfil }}
  \rightline{\hbox to 1.5in{\tenrm  (T/E)\hfil }}     }

\vfill
\leftskip .35in
\parindent=0pt
\baselineskip 16.1pt plus .65pt minus .35pt
\bigskip
\vbox{ \fourteenbf
Relating Physical Observables in QCD without
Scale-Scheme Ambiguity$^\star$
}
\bigskip
\baselineskip 14.1pt plus .65pt minus .35pt
\vbox{\bf  Hung Jung Lu \nostr
{\it
Department of Physics, \nostr
University of Maryland,
College Park, Maryland 20742} }
\smallskip
\vbox{\bf  Stanley J. Brodsky \nostr
{\it   Stanford Linear Accelerator Center,  Stanford, CA 94309}}
\bigskip
PACS numbers: 11.10.Gh, 11.15.Bt, 12.38.Bx, 13.65.+i
\bigskip
\vfill
\parindent=.35in
\baselineskip 24.1pt plus .65pt minus .35pt
\noindent
{\bf Abstract}
\medskip
We discuss the St\"uckelberg-Peterman
extended  renormalization group
equations in perturbative QCD, which express the invariance
of physical observables under renormalization-scale
and scheme-parameter
transformations. We introduce a universal coupling function
that covers all possible choices of scale and scheme.
Any perturbative series in QCD is shown to be equivalent
to a particular point in this function. This function
can be computed from a set of first-order differential
equations involving the extended beta functions.
We propose the use of these evolution equations
instead of perturbative series for numerical evaluation
of physical observables. This formalism is free of
scale-scheme ambiguity and allows
a reliable error analysis of higher-order corrections.
It also provides a precise definition for
$\Lambda_{\overline{\rm MS}}$ as the pole in the
associated 't Hooft scheme.
A concrete application to $R(e^+e^- \to {\rm hadrons})$ is
presented.
\par

\leftskip=0pt
\vfill
\centerline{Submitted to {\it Physical Review D}}
\vfill

\singlespace
\leftline{\us{\hskip 1.5in}}
\vskip -.035in
\hangindent .11in \hangafter 1
\noindent
$^\star$Work supported in part by Department of Energy
contract DE--AC03-76SF00515 (SLAC) and
contract DE-FG05-87ER-40322 (Maryland).
\par }
\endpage


\REF\Stevenson{
            P.~M.~Stevenson,
            Phys.~Lett.~B   100,   61 (1981);
            Phys.~Rev.~D     23, 2916 (1981);
            Nucl.~Phys.~B   203,  472 (1982);
            Nucl.~Phys.~B   231,   65 (1984). }
\REF\Grunberg{
            G.~Grunberg,
            Phys.~Lett.~B    95,   70 (1980);
            Phys.~Lett.~B   110,  501 (1982);
            Phys.~Rev.~D     29, 2315 (1984). }
\REF\BLM{
            S.~J.~Brodsky, G.~P.~Lepage and P.~B.~Mackenzie,
            Phys.~Rev.~D     28,  228 (1983). }
\REF\GorishnyKataevLarin{
            S.~G.~Gorishny, A.~L.~Kataev and S.~A.~Larin,
            Phys.~Lett.~B   259, 144 (1991). }
\REF\SurguladzeSamuel{
            L.~R.~Surguladze and M.~A.~Samuel,
            Phys.~Rev.~Lett. 66, 560 (1991),
            erratum: {\sl ibid.} 66, 2416 (1991). }
\REF\MoreKataev{
            J.~Ch\'yla, A.~L.~Kataev and S.~A.~Larin,
            Phys.~Lett.~B 267, 269 (1991),       \nostr
            J.~Ch\'yla and A.~L.~Kataev,
            preprint CERN-TH-6604/92. }
\REF\TarasovVladimirovZharkov{
            O.~V.~Tarasov, A.~A.~Vladimirov and A.~Yu.~Zharkov,
            Phys.~Lett.~B    93,  429 (1980). }
\REF\StueckelbergPeterman{
            E.~C.~G.St\"uckelberg and A.~Peterman,
            Helv.~Phys.~Acta 26,  499 (1953), \nostr
            A.~Peterman,
            Phys.~Rept.    53~C,  157 (1979).}
\REF\tHooft{
            G.~'t~Hooft in The Whys of Subnuclear Physics,
            Proc. Intern. School of Subnuclear Physics
            (Erice, 1977), ed. A.~Zichichi (Plenum,
            New York, 1979), p. 943. }
\REF\Marshall{
            R.~Marshall,
            Z.~Phys.~C   43, 595 (1989). }
\REF\West{
            G.~B.~West,
            Phys.~Rev.~Lett. 67, 1388 (1991),
            erratum: {\sl ibid.} 67, 3732 (1991). }
\REF\CommentsWest{
            D.~T.~Barclay and C.~J.~Maxwell,
            preprint DTP-91/72 (Durham, UK 1991), \nostr
            M.~A.~Samuel and E.~Steinfeld,
            OSU preprint RN-256, 1992,            \nostr
            L.~S.~Brown and L.~G.~Yaffe,
            Phys.~Rev.~D    45,  398 (1992).}
\REF\ReviewParticleProperties{
            Particle Data Group, Review of Particle Properties,
            Phys.~Rev.~D 45, S1 (1992). }
\REF\OPAL{
            M.~Z.~Akraway {\sl et al.},
            Z.~Phys.~C 49, 375 (1991). }
\REF\Chyla{
            J.~Ch\'yla,
            Phys.~Rev.~D 40, 676 (1989), \nostr
            G.Grunberg,
            {\sl ibid.} 40, 680 (1989).}

\FIG\Figone{
            Pictorial representation of the universal
            coupling function $a(\tau,\{c_i\})$, where
            $\tau$ is the scale parameter and
            $\{c_i\}$ the scheme parameters
            (see definitions in text).
           }
\FIG\Figtwo{
            Graphical representation of the various
            errors involved. For the measurement
            of $\Lambda_\MS^\tH$ (or equivalently
            $\Lambda_R^\tH$) the input experimental
            error must be combined with the scheme
            uncertainty. For the prediction of
            $a_R(\tau)$, the error in $\Lambda_\MS^\tH$
            must be combined with the scheme uncertainty.
           }
\FIG\Figthree{
            Measurement of $\Lambda^\tH_\MS$ from the experimental
            result of $a_R(31.6~\GeV)$. We have parametrized the
            scheme uncertainty with a value $c_3^R = 100$.
            The scheme, experimental and total errors are
            respectively given by
            $\Delta_{\rm sch} = (\tau_5-\tau_3)/2$,
            $\Delta_{\rm exp} = (\tau_6-\tau_2)/2$ and
            $\Delta_{\rm tot} = (\tau_7-\tau_1)/2$.
            There is a one-to-one relationship between
            $\tau$ and $\Lambda^\tH_\MS$ given by
            $\tau=2\beta_0^2\beta_1^{-1}\log
            (31.6~\GeV/1.4443~\Lambda^\tH_\MS)$.
           }
\FIG\Figfour{
            Prediction of $a_\MS(M_Z)$ from the experimental
            result of $a_R(31.6~\GeV)$. By using the extended
            renormalization group equations, the quasi-parallelogram
            $ABCD$ is evolved into the quasi-parallelogram
            $A'B'C'D'$. Notice the inversion of the orientation
            of the parallelograms due to the opposite signs
            of $c_3^R$ and $c_3^\MS$. Notice also the absence
            of scheme uncertainty in the 't Hooft scheme.
           }
\rm

\bigskip

\pageno=1

 The scale-scheme ambiguity problem [\Stevenson,\Grunberg,\BLM]
remains as one of the
major cornerstones impeding precise QCD predictions.
Although all physical predictions in QCD should in principle
be invariant under change of renormalization scale and scheme,
in practice this invariance is only approximate due to
the truncation of their perturbative series.

 Consider the $N$-th order expansion series
of a physical observable $R$ in terms of a coupling constant
$\alpha_S(\mu)$ given in scheme $S$ and at a
scale $\mu$:
$$
R_N = r_0       \alpha_S^p (\mu)
    + r_1 (\mu) \alpha_S^{p+1} (\mu)
    + \cdots
    + r_N (\mu) \alpha_S^{p+N} (\mu) ~.
\eqn\EqR $$
The infinite series $R_\infty$ is renormalization scale-scheme
invariant. However, at any finite order, the scale and
scheme dependencies from the coupling constant $\alpha_S (\mu)$
and from the coefficient functions $r_i (\mu)$ do not
exactly cancel, which leads to a remnant dependence in the
finite series.
Different choices of scale and scheme then lead to
different theoretical predictions.
The availability of next-to-next-to-leading order
results in QCD [\GorishnyKataevLarin,\SurguladzeSamuel]
has accentuated the need for study on the scale-scheme
dependence (e.g., see Ref. [\MoreKataev]).

There have been traditionally two positions on this subject.
The first one is to consider the scale-scheme ambiguity
as intrinsically unavoidable, and interpret
the numerical fluctuations coming from different
scale and scheme choices as the error in the theoretical
prediction. This point of view, aside from being overly
pessimistic, is also very unsatisfactory.
First of all,
in general we do not know how wide a range
the scale and scheme parameters should vary
in order to give a correct error estimate.
Secondly, besides the error due to scale-scheme uncertainties
there is also the error from the omitted higher-order
terms. In such an approach,
it is not clear whether these errors are
independent or correlated.
The error analysis in this context can become quite
arbitrary and unreliable.

A second approach is to optimize the choice of scale
and scheme according to some sensible criteria.
Commonly used scale setting strategies include
the Principle of Minimum Sensitivity [\Stevenson]
(which also optimizes the choice of
scheme), the Fastest Apparent Convergence
criterion [\Grunberg] and the BLM method [\BLM].

In this paper we propose the use of the
Extended Renormalization Group Equations
as a transparent solution to the scale-scheme ambiguity problem.
In this approach, a perturbative series
only serves as an intermediate device
for the identification of scale and scheme parameters.
The ultimate prediction is obtained through
evolution equations in the scale- and scheme-parameter space.
This approach sets the ground for a reliable error analysis
and also provides a precise definition for $\Lambda_\MS$.

We will consider the case of QCD
with $N_f$ massless quarks.
Let us first explain the concept of the universal coupling function
in QCD.
It is well-known that at the renormalization stage
a particular subtraction prescription and
a particular renormalization scale must be specified.
Let us parametrize
the subtraction prescription by an infinite set of continuous
``scheme parameters" $ \{ c_i \} $
and the renormalization scale by $\mu$.
The universal coupling function (see Fig. 1) is the extension of an
ordinary coupling constant to include the dependence
on the scheme parameters $\{ c_i \} $:
$$
  \alpha = \alpha( \mu/\Lambda, \{ c_i \} )  ~.
\eqno\eq $$
For the moment,
the presence of the quantity $\Lambda$
in the previous expression can be justified on
dimensional grounds. We will identify it later with
the 't Hooft scale of the chosen scheme.

Stevenson [\Stevenson] has shown that
one can identify the beta-function coefficients of
a renormalization scheme as its scheme parameters.
That is, if a given scheme has
the following beta-function expansion:
$$
\beta(\alpha)   = {d \over d \log \mu^2}
                  \left( {\alpha \over 4 \pi} \right)
                = -\beta_0
                  \left( {\alpha \over 4 \pi} \right)^2
                  -\beta_1
                  \left( {\alpha \over 4 \pi} \right)^3
                  -\beta_2
                  \left( {\alpha \over 4 \pi} \right)^4
                  + \cdots  ~;
\eqno\eq $$
then, the coefficients $\{ \beta_i ,\ \ i=2,3,\dots \}$
can be considered as the corresponding scheme parameters.
(The scheme invariance of the first two coefficients
$\beta_0$ and $\beta_1$ is a well-known fact. It is
important not to confuse
the coefficients $\beta_i$ with the
$\beta_{(i)}$ functions to be introduced later.)
It will be very convenient to use the first two coefficients
of the beta functions to rescale the coupling constant and
the scale parameter $\log \mu^2$.
Let us define the rescaled coupling constant and the
rescaled scale parameter as
$$
 a    = {\beta_1 \over \beta_0} {\alpha \over 4 \pi} ~, \ \ \ \
 \tau = {2 \beta_0^2 \over \beta_1} \log(\mu / \Lambda)  ~.
\eqno\eq $$
Then, the rescaled beta function takes the canonical form:
$$
\beta( a ) = {d a \over d \tau}
           = - a^2 ( 1 + a + c_2 a^2 + c_3 a ^3 + \cdots ) ~,
\eqno\eq $$
with $c_n = \beta_n \beta_0^{n-1} / \beta_1^n $
for $n=2,3,\cdots$.
This rescaling process serves to ``unitarize" the
expansion coefficients. For a well-behaved scheme in QCD,
we would expect its beta-function expansion to roughly resemble
a geometrical series, at least for the first few
coefficients. In fact, for the $\MS$ scheme we have
$
 c_2^\MS = \beta_2^\MS \beta_0 / \beta_1^2 ~,
$
where [\TarasovVladimirovZharkov]
$$\eqalign{
  \beta_0 & = 11 - {2\over 3} N_f    ~, \cr
  \beta_1 & = 102 - {38 \over 3} N_f ~, \cr
  \beta_2^\MS & =
    {2857\over 2} - {5033\over 18} N_f + {325\over 54} N_f^2 ~, \cr
}\eqno\eq $$
and for $N_f=0,1,2,3,4,5,6$ we have respectively
$c_2^\MS= 1.5103, 1.4954, 1.4692, \str
1.4147, 1.2851, 0.92766, -0.33654$. We can clearly see that indeed
$c_2^\MS$ is of order of magnitude unity.
This should be contrasted with the large value of
$\beta_2^\MS$ before the rescaling process.

The Extended Renormalization Group Equations simply express
the invariance of physical quantities under
scale- and scheme-parameter transformations.
These equations have been studied long ago by
St\"uckelberg and Peterman [\StueckelbergPeterman]
and later by Stevenson [\Stevenson]:
$$\eqalign{
{\d R \over \d \tau} &= \beta {\partial R \over \partial a}
                      + {\partial R \over \partial \tau}  = 0 ~,
                     \cr
{\d R \over \d c_n} &= \beta_{(n)} {\partial R \over \partial a}
                      + {\partial R \over \partial c_n}   = 0 ~.
                     \cr
}\eqn\ExtRenGroEqns $$
Various quantities in these equations need explanation.

First of all, notice the distinction  between
$ \d / \d \tau $ and $ \partial / \partial \tau $.
The partial derivative $\d / \d \tau$
takes into account the full variation
of $R$ under $\tau$ transformation, whereas
the partial derivative $\partial / \partial \tau$
affects only the expansion coefficients of $R$.
An analogous distinction holds between
$\d / \d c_n$ and $\partial / \partial c_n$.
In other words, in the left-hand side
$$ R = R(\tau,\{c_i\}) ~,
\eqno\eq $$
whereas in the right-hand side
$$ R = R( a, r_n( \tau, \{c_i\}) ) ~.
\eqno\eq $$
Mathematically these are two different functions
since their domains are different.
However, it is common practice to use the same notation
for both and draw the difference
only for their partial derivatives.

The fundamental beta function that appears
in Eqs. \ExtRenGroEqns~is defined as:
$$
\beta(a,\{ c_i \} ) \equiv
{\d a \over \d \tau} = -a^2 (1 + a + c_2 a^2 + c_3 a^3 + \cdots ) ~,
\eqno\eq $$
and the extended or scheme-parameter beta functions are
defined as:
$$
\beta_{(n)}(a,\{ c_i \} ) \equiv
{\d a \over \d c_n} ~.
\eqno\eq $$
As shown by Stevenson [\Stevenson],
these extended beta functions can be expressed in
terms of the fundamental beta function. Indeed,
the commutativity of second partial derivatives
$$
  {\d^2 a \over \d \tau \d c_n} = {\d^2 a \over \d c_n \d \tau }
\eqno\eq $$
implies
$$
  {\d \beta_{(n)} \over \d \tau } = {\d \beta \over \d c_n} ~,
\eqno\eq $$
$$
  \beta \beta_{(n)}' = \beta_{(n)} \beta' - a^{n+2} ~,
\eqno\eq $$
where $\beta_{(n)}' = {\partial \beta_{(n)}/ \partial a}$
and $\beta' = {\partial \beta/ \partial a}$.
{}From here
$$\eqalign{
 \beta^{-2} \left( {\beta_{(n)} \over \beta }
            \right)'                              &=
  -a^{n+2}  ~,                                    \cr
  \beta_{(n)} ( a, \{ c_i \} )                    &=
 -\beta   ( a, \{ c_i \} ) \int_0^a dx
  { x^{n+2} \over
    \beta^2 ( x, \{ c_i \} )
  } ~,                                            \cr
}\eqno\eq $$
where the lower limit of the integral has been set
to satisfy the boundary condition
$$
\beta_{(n)} \sim O(a^{n+1}) ~.
\eqno\eq $$
That is, a change in the scheme parameter $c_n$ can
only affect terms of order $a^{n+1}$ and higher in the
evolution of the universal coupling function [\Stevenson].

The extended renormalization group equations in \ExtRenGroEqns~
can be put into the form:
$$\eqalign{
  { \partial R \over \partial \tau }          &=
  - \beta { \partial R \over \partial a }~,   \cr
  { \partial R \over \partial c_n  }          &=
  - \beta_{(n)} { \partial R \over \partial a }~.  \cr
}\eqno\eq $$
These equations can now be interpreted in the following
manner. The left-hand sides represent the variation
of the expansion coefficients of $R$ under
scale-scheme transformations.
Thus, a given perturbative series can be evolved
into another perturbative series, provided we know
the extended beta functions on the right-hand side.
In fact, in the expansion series of $R$ as given
in Eq. \EqR, the only quantities that cannot be modified
under scale-scheme transformations
are the tree-level coefficient $r_0$ and exponent $p$,
since they are renormalization scale- and scheme-independent.

We can standardize a perturbative series
$R=r_0 a^p + \cdots$ by defining
an effective charge $a_R$
(see Grunberg in Ref. [\Grunberg]) by:
$$
 a_R \equiv \left( {R \over r_0}
            \right)^{1/p} ~.
\eqno\eq $$
Since $R$, $r_0$ and $p$ are all renormalization scale
and scheme invariant, the effective
charge $a_R$ is also scale and scheme invariant.
We only need to study the evolution of one effective charge
to another. The appropriate values of $r_0$ and $p$
can always be put back at the end of the analysis.

More concretely, given two perturbative series:
$$\eqalign{
R &= r_0 a_R^p =
 r_0 a^p + r_1 a^{p+1} + \cdots   ~,            \cr
R' &= {r'}_0 a_{R'}^{p'} =
 {r'}_0 a^{p'} + {r'}_1 a^{p'+1} + \cdots ~, \cr
}\eqno\eq $$
we can evolve $R$ into $R'$ by following the next three steps
\item{1)} Obtain the effective charge of $R$
$$
   a_R = \left( {R \over r_0}
            \right)^{1/p}
    = a + { r_1 \over p ~ r_0 } a^2
        + \left( {r_2 \over p ~ r_0} +
                 {1-p \over 2 ~ p^2} ~ {r_1^2 \over r_0^2}
          \right) a^3
        + \cdots ~.
\eqno\eq $$
\item{2)} By using the extended renormalization group equations,
we can change the perturbative coefficients of
$a_R$ and evolve it into
$$
   a_{R'} = a + { {r'}_1 \over p' ~{r'}_0 } a^2
        + \left( {{r'}_2 \over p' ~{r'}_0} +
                 {1-p' \over 2 ~ {p'}^2} {{r'}_1^2 \over {r'}_0^2}
          \right) a^3
        + \cdots ~.
\eqno\eq $$
\item{3)} Obtain $R'$ from $a_{R'}$  by
$$
  R' = {r'}_0 a_{R'}^{p'}  ~.
\eqno\eq $$

Hence, by using the extended renormalization group equations
and by appropriately raising the power and rescaling our
final result, we can evolve a given perturbative series $R$
into any other perturbative series $R'$.
The commutativity of second partial derivatives guarantees
that the final result is independent of the path chosen
for evolution.

Since the universal coupling functions $a(\tau,\{ c_i \} )$
covers all possible choices of scale and scheme,
any effective charge $a_R$ can be expressed in terms
of it:
$$
  a_R = a(\tau_R , \{ c_i^R \} ) ~,
\eqno\eq $$
where $\tau_R$ and $c_i^R$ are respectively the scale
and scheme parameters of $R$.
Notice that this is also true for multiple-scale
processes: given the perturbative series of
a multiple-scale process, we can also define
an effective charge associated to it, and this
effective charge can then be written in terms
of the universal coupling function at a
particular set of values of
$\tau$ and $\{ c_i^R \}$.

The universal coupling function
is dictated by the evolution equations:
$$\eqalign{
    { \d a \over \d \tau }     &= \beta (a,\{ c_i \} )
                                = -a^2 (1+a+c_2 a^2 + \cdots)~,
                               \cr
    { \d a \over \d c_n  }     &= \beta_{(n)} (a,\{ c_i \} )
                                = -\beta (a,\{ c_i \} )
                                  \int_0^a d x
        { x^{n+2} \over \beta^2(x,\{ c_i \} ) }~.\cr
}\eqn\EvolUnivCoup $$
We shall define $a(\tau,\{c_i\})$ here with the boundary condition:
$$
  a(0,\{0\}) = \infty.
\eqno\eq $$

Notice that the above equations contain no explicit
reference to QCD parameters such as the numbers of
colors or the number of flavors. Therefore,
aside from its infinite dimensional character,
$a(\tau,\{ c_i \} )$ is just
a mathematical function like, say, Bessel
functions or any other special function.
Truncation of the fundamental beta function
simply corresponds to evaluating
$a(\tau,\{ c_i \})$ in a subspace
where higher order $c_i$ are zero.
In principle, this function can be computed to
arbitrary degree of precision, limited only
by the truncation of the fundamental beta function.
\footnote*{The $n!$ divergence expected for the beta function
coefficients of physical effective charges
can impose some theoretical limit to the
achievable precision. Further discussion in this
direction is beyond the scope of this paper.}

Notice that the 't Hooft scheme [\tHooft]
defined by
$
  a_\tH(\tau) \equiv a(\tau,\{0\})
$
is privileged since it is
totally devoid of higher-order corrections.
In fact, $a_\tH(\tau)$ is exactly given by
the solution of
$$
  {1 \over a_\tH} + \log
  \left( {a_\tH \over 1+a_\tH} \right) = \tau ~.
\eqno\eq $$

For any single-scale process $R(\mu)$ there exists a scale
$\mu=\Lambda^\tH_R$ for which the scale parameter
$\tau_R = 2 \beta_0^2 \beta_1^{-1}
\log(\mu / \Lambda^\tH_R)$ vanishes.
We will call $\Lambda^\tH_R$ the 't Hooft scale\footnote\dag{For
multiple-scale processes, the sub-manifold
where the scale parameter vanishes defines the
``'t Hooft surface".}
of the $R$-scheme.
To understand the meaning of the 't Hooft scale, let us
consider the $\MS$ scheme coupling constant:
$$
   a_\MS(\mu) = a \left( {2 \beta_0^2 \over \beta_1}
           \log(\mu/\Lambda^\tH_\MS), \{ c_i^\MS \} \right)~.
\eqno\eq $$
Notice that {\sl a priori} we do not know the
behavior of $a_\MS(\mu)$ at $\mu=\Lambda^\tH_\MS$:
it could be infinite, finite, or simply not
well-defined. However, $\mu=\Lambda^\tH_\MS$
is the pole in the 't Hooft scheme associated\footnote\star{There
are infinite 't Hooft schemes, differing only by the
value of the 't Hooft scale
$\Lambda^\tH$. The word ``associated" here means
we are choosing the particular 't Hooft scheme that
shares the same 't Hooft scale with the $\MS$ scheme:
$\Lambda^\tH=\Lambda^\tH_\MS$.}
to the $\MS$ scheme:
$$
   a_{\tH-\MS}(\mu) \equiv a \left( {2 \beta_0^2 \over \beta_1}
           \log(\mu/\Lambda^\tH_\MS),\{ 0 \} \right)~,
\eqno\eq $$
because $a(0,\{ 0 \} )=\infty$ by boundary condition.
Since the 't Hooft scheme is completely free of
higher-order corrections, this provides a precise
definition for $\Lambda_\MS$. This cures the well-known
arbitrariness in the definition of $\Lambda_\MS$
due to the presence of higher order corrections in
the $\MS$ scheme.

Given two effective charges, say
$$\eqalign{
   a_\MS(\mu) & = a \left( {2 \beta_0^2 \over \beta_1}
                \log(\mu/\Lambda^\tH_\MS),\{c_i^\MS\} \right)~,
              \cr
   a_R  (Q) & = a \left( {2 \beta_0^2 \over \beta_1}
                \log(Q/\Lambda^\tH_R),\{c_i^R\} \right)~,
              \cr
}\eqno\eq $$
we can expand $a_R(Q)$ in a power series of
$a_\MS(\mu)$. To this end, we can perform
a Taylor expansion around
the point
$$
(\tau, \{ c_i \}) =
                \left( {2 \beta_0^2 \over \beta_1}
                \log(\mu/\Lambda^\tH_\MS),\{c_i^\MS\} \right)~,
\eqn\ExpansionPoint $$
and generate a series representation for $a_R(Q)$:
$$\eqalign{
 a_R(Q) & =
            a \left( {2 \beta_0^2 \over \beta_1}
              \log(\mu/\Lambda^\tH_\MS) + \bar \tau
              ,\{ c_i^\MS + \bar{c}_i \} \right)~
        \cr
       & = a + \left( { \d a \over \d \tau } \right) \bar \tau
         + \left( { \d a \over \d c_n } \right) \bar{c}_n
       \cr
 & + {1\over 2!} \left[
                    \left(
                       { \d^2 a \over \d \tau^2 }
                    \right) \bar{\tau}^2
                + 2 \left(
                       { \d^2 a \over \d \tau \d c_n }
                    \right) \bar\tau \bar{c}_n
                 +  \left(
                       { \d^2 a \over \d c_n \d c_m }
                    \right) \bar{c}_n \bar{c}_m
                  \right]
 \cr
 & + {1\over 3!} \left[
                    \left(
                       { \d^3 a \over \d \tau^3 }
                    \right) \bar{\tau}^3
                    + \cdots
                 \right]
                + \cdots  ~,
 \cr
}\eqno\eq $$
where
$$\eqalign{
\bar \tau & = { 2 \beta_0^2 \over \beta_1 }
            \left[ \log (Q/\Lambda^\tH_R) -
                   \log (\mu / \Lambda^\tH_\MS)
            \right]  ~,
          \cr
\bar{c}_n & = c_n^R - c_n^\MS ~,
          \cr
}\eqn\TauBarCnBar $$
and $a$ and its derivatives are evaluated at the point
specified in Eq. \ExpansionPoint.
To order $a^4$, we only need the following partial
derivatives:
$$\eqalign{
\left( { \d a \over \d \tau } \right) &= \beta
= -a^2 -a^3 - c_2 a^4 + O(a^5) ~,    \cr
\left( { \d a \over \d c_2  } \right) &= \beta_{(2)}
=  a^3                + O(a^5) ~,    \cr
\left( { \d a \over \d c_3  } \right) &= \beta_{(3)}
= {1\over 2} a^4      + O(a^5) ~,    \cr
\left( { \d^2 a \over \d \tau^2 } \right) &=
  2 a^3 + 5 a^4        + O(a^5) ~,    \cr
\left( { \d^2 a \over \d \tau \d c_2 } \right) &=
  -3 a^4               + O(a^5) ~,    \cr
\left( { \d^3 a \over \d \tau^3  } \right) &=
  -6 a^4               + O(a^5) ~.    \cr
}\eqno\eq $$
After grouping all the terms in powers of $a=a_\MS(\mu)$, we obtain:
$$\eqalign{
a_R(Q) &= a - \bar \tau a^2 + \left( \bar{c}_2
                                   - \bar\tau + \bar\tau^2
                              \right) a^3
       \cr
       &+ \left( {1\over2}\bar{c}_3
               - (c_2+3 \bar{c}_2) \bar\tau
               + {5\over2}\bar\tau^2 - \bar\tau^3
          \right) a^4
       \cr
       &+ O(a^5) ~,
       \cr
}\eqn\UniversalExpansion $$
where $\bar\tau$ and $\bar{c}_n$ are as given in
Eq. \TauBarCnBar, and
$c_2=c_2^\MS$.
Although we have used $R$ and $\MS$ schemes in the derivation
of this last formula, naturally it is also valid for any
other pair of effective charges. Notice the occurrence
of $\bar\tau$ and $\bar{c}_i$ in all higher order
coefficients. By using the evolution equations
in \EvolUnivCoup, we are effectively performing
a partial resummation of the perturbative series to
all orders.

Any physical quantity  $R$ calculated
perturbatively in the $\MS$ scheme can
be put into the above standard form. Hence,
we can use Eq. \UniversalExpansion \
to identify the scale and scheme parameters of
the effective charge of $R$.
In other words, to give the numerical prediction
for $R=r_0 a^p + \cdots$ we proceed to
\item{1)} compute its perturbative series in terms
of the coupling constant in some scheme, say $\MS$ scheme,
\item{2)} standardize the series and identify the
scale and scheme parameters $(\tau_R, \{c_i^R\})$
order-by-order
via Eq. \UniversalExpansion,
\item{3)} from the knowledge of $\Lambda_\MS^\tH$,
evolve the universal coupling function
$a(\tau, \{c_i\})$ to the point $(\tau_R, \{c_i^R\})$.
Put back the values of $r_0$ and $p$ if necessary.

Naturally we can also go in the opposite direction and
obtain $\Lambda_\MS^\tH$ from the experimental
measurement of $R$.
In Fig. 2 we show the various experimental and
theoretical errors involved in the analysis.
For the measurement of $\Lambda_\MS^\tH$, the
input experimental error must be combined with
the scheme uncertainty to give the error estimate
for $\Lambda_\MS^\tH$. Similarly, for the prediction
of $a_R(\tau)$ the error from $\Lambda_\MS^\tH$
must be combined with the scheme uncertainty
in order to give the prediction error.

As a concrete example, let us consider the total hadronic
cross section in $e^+e^-$ annihilation
$R(Q) = R(e^+e^- \to {\rm hadrons})$ recently
calculated to order $\alpha^3$
[\GorishnyKataevLarin,\SurguladzeSamuel].
For five light-quark flavors we have
$$\eqalign{
  R(Q) &= {11\over3}
         \left[ 1 + {\alpha \over \pi}
                  + 1.4092 \left( {\alpha \over \pi} \right)^2
                 - 12.8046 \left( {\alpha \over \pi} \right)^3
         \right]
       \cr
       &\equiv {11\over3}
         \left[ 1 + {\alpha_R(Q) \over \pi}
         \right] ~,
       \cr
}\eqno\eq $$
where we have used
$ \alpha = \alpha_\MS (Q) $
for the strong coupling constant.

By putting $R(Q)$ into the standard form of Eq. \UniversalExpansion:
$$
  a_R(Q) = a + 1.1176~a^2 - 8.05426~a^3 ~,
\eqno\eq $$
where $a=a_\MS(Q)$, we can identify
$$
   \bar \tau   = -1.1176 ~,    ~~~~~~~
   \bar{c}_2 = -10.4209~.
\eqno\eq $$
Knowing that $\bar\tau = 2 \beta_0^2 \beta_1^{-1}
\log(\Lambda^\tH_\MS / \Lambda^\tH_R)$
and
$\bar c_2 = c_2^R - c_2^\MS$, we conclude that
$$
   \Lambda^\tH_R = 1.4443~ \Lambda^\tH_\MS ~,    ~~~~~~~
   c_2^R = -9.4932~.
\eqno\eq $$

Experimentally [\Marshall]
we have
$$
  r(31.6~\GeV) = {3\over 11} R(31.6~\GeV) = 1.0527 \pm 0.0050  ~,
\eqno\eq $$
which gives
$$
   a_R(31.6~\GeV) = 0.0665 \pm 0.0063 ~.
\eqn\ExpR $$

We have to take into account the scheme uncertainty
in addition to the experimental error in order
to quote a correct error estimate for $\Lambda^\tH_\MS$
(see Fig. 3).
The scheme uncertainty of $R$ can be quantified
by a reasonable estimate of its next scheme parameter:
$c_3^R$. G.~B. West [\West] has put an estimate $r_4=-158.6$
for the coefficient $r_4$ in
$$
  R = 3 \sum Q_f^2
      \left[ 1
          +  r_1 \left({\alpha\over\pi}\right)
          +  r_2 \left({\alpha\over\pi}\right)^2
          +  r_3 \left({\alpha\over\pi}\right)^3
          +  r_4 \left({\alpha\over\pi}\right)^4
          +  \cdots
      \right] ~.
\eqno\eq $$
By reducing this equation into the standard form of
Eq. \UniversalExpansion, we arrive to a value
$\bar c_3 = c_3^R - c_3^\MS = -99.474$.
Assuming $c_3^\MS$ is of order unity,
we conclude that
$\left| c_3^R \right| \sim 100$.
Although we have some reservation on West's estimate
(see Ref. [\CommentsWest]), we shall
nonetheless use it to illustrate our procedure. A
better estimate of $c_3^R$ will lead to a better
error estimate for $\Lambda^\tH_\MS$.

In Fig. 3 we show the universal charge for
$a_0(\tau) = a(\tau,\{c_2=c_2^R, c_3=c_4=\dots=0 \})$
and its evolution under a scheme uncertainty $c_3=\pm 100$ to
$a_\pm(\tau) = a(\tau,\{c_2=c_2^R, c_3=\pm 100,
c_4=c_5=\dots=0 \})$.

The evolution in the scheme parameters are dictated by:
$$\eqalign{
  {\d a \over \d c_2}
  & =  \beta_{(2)} = -\beta \int_0^a dx { x^4 \over \beta^2 }
       \sim a^3 + O(a^5) ~,
  \cr
  {\d a \over \d c_3}
  & =  \beta_{(3)} = -\beta \int_0^a dx { x^5 \over \beta^2 }
       \sim {1\over 2} a^4 + O(a^5) ~,
  \cr
}\eqn\ApproxEvol $$
For our region of interest ($a \sim 0.07$) the first
term in each expansion series suffices.
But we should use the full integro-differential
equation whenever we want to evolve $a$
to a higher-value region. This will not only
improve the accuracy of our result, but also
will respect the commutativity of the
second-order partial derivatives of $a$
and thus ensure the independence of the
result on the choice of integration path.

To obtain $a_0$ and $a_\pm$ defined
above, we follow the next steps:
\item{1.-} Generate the 't Hooft scheme coupling $a_\tH$
by solving iteratively
$$
  a_\tH={1\over \tau + \log
       \left( 1 + 1/a_\tH
       \right)
      } ~.
\eqn\atHooft $$
\item{2.-} Evolve $a_\tH$ to $a_0$ by displacing
in $c_2$. From \ApproxEvol~ we have:
$$
  a_0={ a_\tH \over
       \left( 1 - 2~ c_2^R~ {a_\tH}^2
       \right)^{1/2}
      } ~.
\eqn\azero $$
\item{3.-} Evolve $a_0$ to $a_\pm$ by displacing
in $c_3$. From \ApproxEvol~ we have:
$$
  a_\pm={a_0 \over
       \left( 1 \mp {3\over2}~ c_3^R~ a_0^3
       \right)^{1/3}
       } ~.
\eqn\apm $$

In Fig. 3 we show the various errors involved in this analysis.
Numerically we find the experimental, scheme, and total errors
for $\tau$ to be:
$$\eqalign{
\Delta \tau_{\rm exp} &= (\tau_6-\tau_2)/2 = 1.41 ~, \cr
\Delta \tau_{\rm sch} &= (\tau_5-\tau_3)/2 = 0.22 ~, \cr
\Delta \tau_{\rm tot} &= (\tau_7-\tau_1)/2 = 1.63 ~. \cr
}\eqno\eq $$
These errors can be translated into uncertainties
in $\Lambda^\tH_\MS$
since there is a one-to-one correspondence
between $\tau$ and $\Lambda^\tH_\MS$.
We can see that most error comes from the experimental error in $a_R$.
We can also see that the experimental error and the scheme error
are highly uncorrelated since
$\Delta \tau_{\rm tot} \sim \Delta \tau_{\rm exp} + \Delta \tau_{sch}$.
Numerically we obtain $\tau_1=10.129, \tau_4=11.666$ and $\tau_7=13.379$.
Knowing that
$$
  \tau = {2 \beta_0^2 \over \beta_1} \log
         \left( { 31.6 ~\GeV \over 1.4443 ~\Lambda^\tH_\MS }
         \right) ~,
\eqno\eq $$
we arrive at the following result for $\Lambda^\tH_\MS$:
$$
  \Lambda^\tH_\MS = 470^{+310}_{-200} ~\MeV  ~.
\eqno\eq $$
If there were no experimental error, the
estimated scheme uncertainty would lead to
$\Lambda^\tH_\MS = 470^{+40}_{-30} ~ \MeV  ~$.\footnote\star{We
have rounded off the above results. More precise values are
$ \Lambda^\tH_\MS = 472^{+310}_{-204} ~\MeV  $ for the
case including experimental error, and
$\Lambda^\tH_\MS = 472^{+35}_{-33} ~ \MeV  ~$ for the
case without experimental error.}

As a second application of our formalism, we will show
next how to use the experimental result of $a_R(31.6 ~\GeV)$
to predict other effective charges. Specifically, we
will give a prediction for $a_\MS(M_Z)$, where
$M_Z=91.173 ~\GeV$ is the mass of the $Z$-boson.
The evolution
of $a_R(31.6~\GeV)$ to $a_\MS(M_Z)$ is illustrated
in Fig. 4. Notice that the experimental and
the scheme uncertainties confine the correct
result for $a_R$ into an approximate
parallelogram $ABCD$.
We can then evolve this parallelogram
into any other scheme and scale.
We will use
$c_3^R=\pm 100$ and $c_3^\MS= \pm 1$ to estimate
the scheme uncertainties in $a_R$ and $a_\MS$.
For $a_\MS(M_Z)$, the parallelogram $ABCD$
is evolved into the parallelogram $A'B'C'D'$.
Notice the inversion of the orientation of the
new parallelogram due to the opposite signs
of $c_2^R$ and $c_2^\MS$.
Notice also the absence of scheme uncertainty
in the 't Hooft scheme.

{}From $\tau_A = 10.129$ and $\tau_C=13.379$ and
knowing that $\tau_\MS = \tau_R - \bar \tau$
with $\bar\tau = 2 \beta_0^2 \beta_1^{-1}
\left[ \log(M_Z/\Lambda_\MS) -
       \log(31.6 ~\GeV/\Lambda_R)
\right]=4.339$,
we find $\tau_{A'}=14.468$ and $\tau_{C'}=17.718$.
{}From here and using $c_2^\MS=0.92766$ and
$c_3^\MS=1$ in the $\MS$ version of the
Eqs. \azero ~and \apm, we obtain:
$a^{A'} = 0.05772$ and
$a^{C'} = 0.04818$.    Hence, we arrive at the prediction
$$
  a_\MS(M_Z) = 0.0530 \pm 0.0048 ~,
\eqno\eq $$
or equivalently,\footnote\dag{This
value is higher than the world average
$\alpha_\MS(M_Z) = 0.1134\pm0.0035$
quoted in the Review of Particle Properties
[\ReviewParticleProperties] but still consistent
with other quoted values for $\alpha_\MS(M_Z)$.
For instance,
$\alpha_\MS(M_Z) = 0.118\pm0.008$ is
obtained by OPAL [\OPAL]. The detailed analysis
of consistency between the various experimental
results is beyond the purpose of this paper.
}
$$
  \alpha_\MS(M_Z) = 0.132 \pm 0.012 ~.
\eqno\eq $$

Let us show next that our formalism is closely related to
the FAC (Fastest Apparent Convergence) criterion when
only the next-to-leading-order coefficient is known.
We define FAC here as the condition of a vanishing
next-to-leading-order coefficient.

Given
$$
a_R(Q) = a_\MS(\mu) - \bar \tau a_\MS^2(\mu) ~,
\eqn\NLO $$
with
$$
\bar \tau = 2 \beta_0^2 \beta_1^{-1}
\left[ \log (Q/\Lambda^\tH_R) - \log(\mu/\Lambda^\tH_\MS)
\right]  ~,
\eqno\eq $$
and assuming a complete lack of knowledge of
the scheme parameters $\{c_i^\MS\}$ and $\{c_i^R\}$,
we cannot do much better than to approximate:
$$\eqalign{
a_R(Q) &= a \left( {2 \beta_0^2 \over \beta_1} \log (Q/\Lambda^\tH_R),
                   \{ 0 \}
            \right)  ~,
      \cr
a_\MS(\mu) &= a \left( {2 \beta_0^2 \over \beta_1}
                       \log (\mu/\Lambda^\tH_\MS),  \{ 0 \}
                \right) ~,
      \cr
}\eqno\eq $$
and absorb the uncertainty  from scheme-parameters
into our theoretical error.
However, the last two equations imply that:
$$
  a_R(Q) = a_\MS \left( \mu= {\Lambda^\tH_\MS \over \Lambda^\tH_R} Q
                 \right) ~.
\eqno\eq $$
Hence, by setting the coefficient $\bar \tau$  in Eq. \NLO  \
to zero, we obtain the correct ratio for the two scales.
Actually, this holds true in general: by applying the
FAC criterion, we always obtain the correct ratio of the
't Hooft scales, whether or not we know the
scheme parameters. Thus, despite its
apparent na\"{\i}veness, FAC actually constitutes a
correct first step towards the elimination of scale-scheme
ambiguity.

Finally let us comment on the definition of effective charges.
The definition of $\alpha_R$ by $R\equiv r_0 \alpha_R^p$ is not
the only possibility [\Chyla]. In fact, we can apply the
extended renormalization group technique to the effective
BLM charge [\BLM]
in the following manner. Given two physical
quantities $R$ and $R'$ computed in a particular scheme
$$\eqalign{
 R(Q)   &= r_0(Q) \alpha^p(\mu)
        + \left( r_{10}(Q) + r_{11}(Q,\mu) \beta_0
          \right) \alpha^{p+1}(\mu) + \cdots ~,
        \cr
 R'(Q')   &= r'_0(Q') \alpha^{p'}(\mu')
        + \left( r'_{10}(Q') + r'_{11}(Q',\mu') \beta_0
          \right) \alpha^{p'+1}(\mu') + \cdots ~,
        \cr
}\eqno\eq $$
where $\beta_0=11- {2\over 3} N_f$ is the first
beta function coefficient,
we can define their ``effective BLM charges" by
$$\eqalign{
 R(Q)   &= r_0(Q) \alpha_{R-BLM}^p(Q)
        +  r_{10}(Q)
           \alpha_{R-BLM}^{p+1}(Q) + \cdots ~,
        \cr
 R'(Q')   &= r'_0(Q') \alpha_{R'-BLM}^{p'}(Q')
        +  r'_{10}(Q')
           \alpha_{R'-BLM}^{p'+1}(Q') + \cdots ~,
        \cr
}\eqno\eq $$
with
$\alpha_{R-BLM}(Q)=\alpha(\mu^\star)$ where
$\mu^\star$ is the solution of $r_{11}(Q,\mu^\star)=0$, and similarly
$\alpha_{R'-BLM}(Q')=\alpha(\mu'^\star)$ where
$\mu'^\star$ is the solution of $r'_{11}(Q',\mu'^\star)=0$.
With this choice of scale, vacuum polarization contributions
are associated with the charge rather than the expansion coefficients,
and the scale tends to reflect the mass of the virtual gluons.
We can then apply the evolution equations to
$\alpha_{R-BLM}(Q)$ and evolve it to
$\alpha_{R'-BLM}(Q')$.
Alternative definitions of effective charges have also been
discussed in Ref. [\Chyla]. However, as we have shown in this paper
any convenient choice of effective charge can be used
to relate physical observables. In  practice
we can adhere to the definition
$R\equiv r_0 \alpha_R^p$ which has the advantage that
$r_0$ and $p$ are renormalization scale-scheme
invariant quantities.

To summarize, we have explained the use of extended
renormalization group equations to relate physical
observables.
The most distinctive feature of this formalism is
that, in this approach,
the perturbative series of a physical observable
only serves to identify the scale and scheme parameters.
The final prediction is obtained by the
evolution of a universal coupling function.
The prediction is scale-scheme independent
in the sense that given the initial perturbative series
in any scheme at any scale, we will always obtain its
correct scale and scheme parameter and hence arrive
at the same prediction.
We have shown that
this formalism sets the ground for a reliable error
analysis, and that $\Lambda_\MS$ can be unambiguously
defined as the pole in the associated 't Hooft scheme.
Finally, we have shown that
this formalism is equivalent to the fastest apparent convergence
criterion in the absence of information on scheme parameters.

\vskip 0.5in
\centerline{\bf ACKNOWLEDGEMENTS}
\vskip 0.5in

This work was supported by
Department of Energy contracts DE--AC03- \str
76SF00515 and DE--FG05--87ER--40322.

\refout\endpage
\figout\endpage
\bye